\documentclass[useAMS,usenatbib]{mn2e}
\pdfoutput=1
\usepackage{amsmath}
\usepackage{mathtext,amssymb,amsmath}
\usepackage{epsfig}
\usepackage{graphics}

\renewcommand{\vector}[1]{\ensuremath{\mathbf{\boldsymbol #1}}}

\newcommand{\pardirone}[2]{\dfrac{\partial  #2}{\partial #1}}

\title[The accretion flow inside the last
  stable orbit]{The thickness of a weakly-magnetized accretion flow inside the last
  stable orbit of a Kerr black hole} 

\author[Abolmasov]{P. Abolmasov\thanks{E-mail: pavel.abolmasov@gmail.com}\\
Sternberg Astronomical Institute, Moscow State University,
  Moscow, Russia 119992\\}

\begin{document}

\date{Accepted ---. Received ---; in
  original form --- }

\pagerange{\pageref{firstpage}--\pageref{lastpage}} \pubyear{2014}

\maketitle

\label{firstpage}

\begin{abstract}
If accretion disc contains weak frozen-in entangled magnetic fields, their
dynamical effect may be important inside the last stable orbit because of the
decompression near the sonic point. Here, I consider the radial
and vertical structure of a nearly
free-falling flow inside the last stable orbit of
a thin disc around a Kerr black hole. The thickness of such a flow is
determined primarily by the vertical stress created by radial and azimuthal  
magnetic fields. The thickness
is predicted to oscillate vertically around its equilibrium value determined
by the magnetic field balance with gravity. For thin discs,
this thickness is much larger than that of the accretion disc itself. 
Numerical simulations
with {\it HARM2d} show the vertical structure is more complicated. In
particular, magnetically supported disc seems to be unstable to segregation of
matter into thinner streams with the vertical scale determined by thermal
pressure or other processes. 
\end{abstract}


\section{Introduction}

A thin accretion disc around a black hole ends near the innermost stable
circular orbit (ISCO) radius where it becomes possible for the matter to
accrete without angular momentum loss \citep{ST83}.
Classical works on disc accretion usually ignore the matter inside the
ISCO because the accretion proceeds on dynamical rather than viscous time
scales here, and the energy dissipation is relatively low. 

On the other hand, there are reasons, both observational and theoretical,
for this region to become a site of intense dissipation. In particular, the
sizes of X-ray emitting regions in lensed quasars \citep{chartas,
  morgan12,chenxrays} seem to
have sizes comparable to the last stable orbit radius, 
hence their fueling requires
energy dissipation localized very close to the black hole. On the other hand, there are
energy reservoirs that can give away some part of their energy before it is
advected by the black hole. 
Three of them should be mentioned separately:

\begin{itemize}

\item Radiation trapped inside the optically thick disc. Trapped radiation is
  important for large accretion rates (comparable to Eddington) when the disc is hot and its optical
  depth high. The vertical optical depth becomes smaller near the inner edge of the
  disc hence the disc radiation should have an additional component emitted by
  its innermost parts. I am not aware about any papers where this contribution
  was estimated, save for my work \citet{eigen13} where the role of this
  radiation for rotational evolution due to radiation capture was
  considered. 

\item Magnetic fields generated and amplified in the disc. Firstly, magnetic
  fields are efficient in extracting the rotational energy of the black hole
  \citep{BZ77,krolik99}. Secondly, their reconnection and dissipation due to
  general relativity effects, Ohmic dissipation and reconnections 
  may heat the gas inside the ISCO up to
  X-ray temperatures limited primarily by inverse Compton (IC) losses (see,
  for instance, \citet{PK95})
  and by the short free-fall time.

\item Thermal and excess potential energy of the matter connected to disc thickness. This
  energy may be realized through gas heating in shock waves after the strong
  pressure drop near the sonic point. 

\end{itemize}

All these energy sources may have luminosities about $\left(H/r\right)^2\dot{M}c^2$, 
comparable to Eddington if the
mass accretion rate is close to or higher than the critical
(Eddington) accretion rate. 
{ For X-ray binaries, the contribution of the inner flow to the radiation
  emitted by the source was considered by \citet{zhux} who find that up to
  15\% of the luminosity of the source may be emitted from inside the ISCO.
}
Transparency and high temperatures make these
parts of the accretion disc the possible source of X-rays in quasars with the
size comparable to the last stable orbit. Thus, inside the last stable orbit,
there is a region of possible observational importance with the structure
rarely considered in research papers.

\bigskip

Magnetic energy also depends on the unknown magnetization
parameter $\beta = p / p_M$ (total to magnetic pressure). If the viscous
stresses in the disc are connected to more or less isotropic magnetic fields, dimensionless viscosity
parameter is $\alpha \sim B_r B_\varphi / 4\pi p \sim 2\beta^{-1}$. 
It seems a reasonable approximation to
assume $\alpha \beta \gtrsim 1$. 

{ Recent more comprehensive estimates
  involving three-dimensional (3D) MHD modeling argue for a somewhat smaller
  universal value of 
  $\alpha \beta \simeq 0.4\div 0.5$ \citep{sorathia}. This should be borne in
  mind when comparing the simulation results to the real astrophysical
  applications. Smaller value of $\alpha \beta$ also puts slightly 
  stronger limitations upon the approximations used in the present work (see
  section~\ref{sec:disc}). }

The ISCO sets the inner boundary of the region where a thin rotationally supported
disc can exist and
accretion is impossible without dissipation. 
In real discs, this boundary is smeared, and real discs should be transonic
(in sense of radial
velocity) with the slow, viscosity-driven radial creep of the accretion disc
evolving into the supersonic free fall toward the horizon. 
 Supersonic radial motion means that energy and momentum are transferred
  in radial direction primarily by the motion of the falling matter, and the
  matter itself moves nearly along the geodesics conserving its energy and angular
  momentum. Using the term ``free fall'', I refer to idealized inward
  motion along geodesics or to motion that is considered as such. 

The standard accretion disc model \citep{SS73} is also known as the thin
disc. Its vertical structure may be decoupled from the structure in radial
direction. All the vertical motions are slow, and the vertical structure of
the disc is determined by hydrostatic equilibrium. If the flow inside the
last stable orbit preserves its disc-like geometry (its thickness is smaller
than its radial extension), this approximation should probably hold. However,
the vertical equilibrium needs some time to establish that implies existence
of a transition region near the edge of the disc.

The structure of the transonic gas-dynamic flow near the ISCO 
was thoroughly considered by
\citet{betch05}. During the free-fall stage, the gas streamlines may intersect
each other in one or several points (nodes) dependent on the black hole rotation parameter and
initial disc thickness. 
A crucial feature of the transonic flow inside the ISCO is strong
decompression due to the increase in radial velocity. Surface density changes by
a factor of $\sim v_r^{disc}/v_K \sim \alpha (H/r)^2 \ll 1$, where $H$ is disc
thickness at radial distance $R$. 
Gas pressure decreases even stronger, at least as $\sim \left(\alpha
(H/r)^2\right)^\Gamma$, where $\Gamma \sim 5/3$ is the adiabatic
index. Moreover, the pressure in the inner parts of accretion discs often
includes important contribution from radiation pressure that drops when the flow
becomes optically thin\footnote{Scattering vertical optical depth of a free-falling flow
  is of the order of mass accretion rate in Eddington units,
$\tau \simeq \varkappa_T \Sigma \sim \dfrac{\varkappa_T \dot{M}}{4\pi R u^r}
  \sim \dfrac{\dot{M}}{\dot{M}_{Edd}}$, where
  $\varkappa_T$ is Thomson opacity, $\Sigma$ is surface density,
  $\dot{M}_{Edd} = \dfrac{4\pi GM}{c\varkappa_T}$ is the Eddington (critical)
  mass accretion rate. Hence unless
  accretion is significantly supercritical \citep{SS73}, the inner flow has optical depth
  less than several.}. All this
implies that, unless the disc scale height changes by a considerable amount,
the vertical structure of the disc is very far from equilibrium.
 
One pressure source that survives this density drop is the
radial magnetic field. The other two components decrease linearly with the
density drop but the radial component of magnetic field changes continuously
due to flux conservation (approximately as $B^r \propto (H R)^{-1}$). If the magnetic fields are in equipartition inside
the accretion disc, Maxwellian stresses $\sim B^2$ will dominate over other
pressure contributions inside the ISCO.
This makes the tangential (vertical) stress the primary force to balance the
gravity in the free-falling region inside a thin disc and thus to influence its
vertical structure. Depending on the parameter values (Kerr parameter, disc
thickness and radial velocity in the disc), the vertical structure may be closer either
to the free-fall solution (as in \citet{betch05}) or to a magnetically
supported analogue of the thin accretion disc. 

Apart from the radial, azimuthal magnetic field component plays important
role because the magnetic field is wound up by differential rotation. 
As it will be shown in section~\ref{sec:mag}, the azimuthal magnetic fields
easily become stronger than radial. 
If both $B^r$ and $B^\varphi$ are present, magnetic stress may also affect
momentum transfer in radial direction. This effect is ignored here but may
become important if either the matter is strongly magnetized ($\alpha\beta \gg
1$) or the disc thickness is high.
In section~\ref{sec:disc}, I consider the limits of this approximation. 

In this work, I consider the radial and vertical structure of an equatorial 
thin disc-like flow that
has negligible gas pressure but strong enough magnetic fields to affect the
vertical force balance.
Radial structure is assumed to be governed by gravity
only, $u^z \ll u^{r} \sim u^{\varphi} \sim u^t$. 

The properties of such a flow are derived
from the fundamental conservation laws (see next section). { Radial structure
may be simply considered as a result of energy and angular momentum
conservation. } Vertical structure,
however, requires evaluation of the vertical gravity in the
co-moving frame (see section \ref{sec:az}). The role of magnetic field components
is considered in section \ref{sec:mag}. The Cauchy problem for the vertical
scale height is formulated 
and the results are presented in section \ref{sec:res}. To check my
semi-analytical results, I also compute a series of numerical two-dimensional
models using the publicly available code {\it HARM}. Results of these
simulations are given and compared to the simplified model in 
section \ref{sec:harm}. 
Discussion and conclusions are given in section \ref{sec:disc}. 

\section{Ideal flow inside the last stable orbit}

\subsection{Basic assumptions}\label{sec:basic}

Let us consider the supersonic part of a thin transonic equatorial flow upon a Kerr black
hole. Radial motion is practically a free fall, but for the vertical
structure, both gravity and pressure gradients will be considered. 

 I consider Kerr metric in 
Boyer-Lindquist coordinates \citep{BLcoord} $t$, $r$, $\varphi$,
$\theta$.

$$
ds^2 = -\alpha^2 dt^2 + \frac{\Sigma^2}{\rho^2} \sin^2\theta
\left(d\varphi-\omega_{LT} dt\right)^2 + \frac{\rho^2}{\Delta} dr^2+\rho^2
d\theta^2
$$

Here:

$$
\alpha^2 = \frac{r^2 \Delta}{\Sigma^2}
$$

$$
\rho^2 = r^2+a^2\cos^2\theta
$$

$$
\Sigma^2 = (r^2+a^2)^2-a^2\Delta \cdot \sin^2 \theta
$$

$$
\Delta = r^2-2r+a^2
$$

$$
\omega_{LT} = \frac{2ar}{\Sigma^2}
$$

All the calculations are performed near the equatorial plane ($\cos \theta
\simeq 0$). To calculate the vertical gravity, terms of the order
$\cos^2\theta$ should be retained in metric components. 
Normalized dimensionless units are used $GM=c=1$.

\bigskip

In the most general form, energy-momentum conservation may be written as:

$$
T^{ik}_{;k}=0
$$

Continuity equation:

$$
(\rho u^i)_{;i}=0
$$

Here, energy-momentum tensor $T^{ik} = \rho u^i u^k + p g^{ik} + T_{EM}^{ik}$,
and practically everywhere below the pressure term is dropped ($p=0$) because,
as it was shown in the Introduction, magnetic tensions overwhelm thermal
pressure inside the last stable orbit. The dynamical
equation for a test particle may be thought of as a consequence of the above
two conservation laws:

$$
u^k u^i_{;k} = -\dfrac{1}{\rho} \left(T_{EM}\right)^{ik}_{;k} - \dfrac{1}{\rho}p^{;i}
$$

If one considers the vertical structure:

$$
u^r u^z_{, r} = a_z +g_z,
$$

where $a_z$ is the acceleration created by magnetic tensions (will be
considered in section~\ref{sec:mag}) and thermal pressure, and $g_z$ is
vertical gravity term  (see section~\ref{sec:az}) responsible for the
difference between covariant and partial derivatives in the above expressions.
 To estimate the vertical gravity, Christoffel symbols should be
 expanded at least to second order in $\cos \theta$. 
Together with
$\theta$, vertical distance $z$ is also used, $z = r \cos \theta
\ll r$. For any tensor quantity or Christoffel symbol, coordinate
transformation from $\theta$ to $z$ near equator 
requires multiplication (division) by the factor of
$|dz/d\theta| = r$ if the quantity is contravariant (covariant) in
$\theta$. Note that the corresponding metric component is trivial
$g_{zz}=g^{zz} =1$ and hence for any vector or tensor quantity evaluated near
the equatorial plane upper and lower position of $z$ index are equivalent. 

\subsection{Radial structure}\label{sec:radial}

 For a particle falling free in Kerr space-time, energy $-u_t$ and angular
momentum $u_\varphi$ are conserved. 
Radial velocity can be expressed straightforward through the conserved
4-velocity components:

$$
g^{ik} u_i u_k = -1
$$

\begin{equation}\label{E:ur}
u_r \simeq \sqrt{-g_{rr}(g^{tt}u_t^2+2g^{\varphi t}u_tu_\varphi + g^{\varphi\varphi}u_\varphi^2+1)}
\end{equation}

Contravariant velocity component may be calculated as $u^r = g^{rr} u_r$.

\subsection{Vertical acceleration}\label{sec:az}

Vertical acceleration as measured in the co-moving frame:

$$
a_z = u^k u_{z;k} = u^k \times \left( u_{z,k} - \Gamma^l_{zk}u_l\right) 
$$

Since the flow is one-dimensional (we neglect vertical motions with respect to
radial), this relation may be re-written as:

$$
u_{z,r} = \dfrac{1}{u^r}\left(a_z + \Gamma_{lzk} u^l u^k \right)
$$

where $g_z = \Gamma_{lzk} u^l u^k$ is vertical gravity. 
Motion along a geodesic corresponds to $a_z=0$, while in a more general case
$a_z$ may be connected to some external force. 

$$
\begin{array}{l}
\hspace{-0.5cm}  g_z =  \Gamma_{rzr} \left(g^{rr}\right)^2 u_r^2 + \Gamma^t_{zt}u^t u_t +
\Gamma^\varphi_{zt}u^t u_\varphi + \Gamma^t_{z\varphi}u^\varphi u_t +
\Gamma^\varphi_{z\varphi}u^\varphi u_\varphi = \\
= \Gamma_{rzr} g^{rr} u_r^2 + \left( \Gamma_{tzt} (g^{tt})^2 + 2\Gamma_{\varphi
  z t} g^{\varphi t} g^{tt} + \Gamma_{\varphi z \varphi} (g^{\varphi
  t})^2\right) u_t^2+\\
+\left( \Gamma_{tzt} (g^{\varphi t})^2+ 2 \Gamma_{\varphi z t} g^{\varphi
  t}g^{\varphi \varphi} + \Gamma_{\varphi z \varphi} (g^{\varphi\varphi})^2
\right) u_\varphi^2+\\
+\left( \Gamma_{tzt} g^{\varphi t}g^{tt}+  \Gamma_{\varphi z t} \left( (g^{\varphi
  t})^2+g^{tt}g^{\varphi\varphi}\right) + \Gamma_{\varphi z \varphi}
g^{\varphi\varphi}g^{\varphi t}
\right) u_t u_\varphi \\
\end{array}
$$

If $u^r$ is substituted following (\ref{E:ur}):

$$
\begin{array}{l}
g_z = u_\varphi^2 \left( g^{\varphi\varphi} \Gamma^\varphi_{z \varphi} +
g^{\varphi t} \Gamma^\varphi_{zt}- g^{\varphi\varphi} \Gamma^r_{zr}\right) +\\
\qquad{} +u_\varphi u_t \left( g^{tt} \Gamma^\varphi_{z\varphi} + g^{\varphi
  t} \times \left(\Gamma^\varphi_{z\varphi} + \Gamma^t_{zt} - 2\Gamma^r_{zr}
\right)+ g^{\varphi \varphi} \Gamma^t_{z\varphi}\right) + \\
\qquad{} + u_t^2 \left(g^{tt} \Gamma^t_{zt} + g^{\varphi t}
\Gamma^t_{z\varphi} - g^{tt} \Gamma^r_{zr}\right) - \Gamma^r_{rz}
 \end{array}
$$

Relevant metric components and 
Christoffel symbols evaluated near $z=0$ (leading terms in
$z/r$ retained) are equal: 

$$
g^{rr} = \dfrac{\Delta}{r^2}
$$

$$
g^{tt} = -\dfrac{1}{r\Delta} \left(r^3 + a^2 (r+2) \right)
$$

$$
g^{\varphi\varphi} = \dfrac{r-2}{r\Delta}
$$

$$
g^{\varphi t} = -\dfrac{2a}{r\Delta},
$$

where $\Delta = r^2 + a^2 -2r$.

$$
\Gamma_{rzr} =\dfrac{a^2}{r^2\Delta} z
$$

$$
\Gamma_{\varphi z\varphi} =-\left(1+\dfrac{a^2(r^3+4r^2+2a^2)}{r^5}\right) z
$$

$$
\Gamma_{\varphi zt} =\dfrac{2a (r^2+a^2)}{r^5} z
$$

$$
\Gamma_{tzt} =-\dfrac{2a^2}{r^5} z
$$

$$
\Gamma^r_{zr} = \dfrac{a^2}{r^4} z
$$

$$
\Gamma^\varphi_{z\varphi} = -\dfrac{r^3+2a^2}{r^5} z
$$

$$
\Gamma^t_{zt} = \dfrac{2a^2}{r^5} z
$$

$$
\Gamma^\varphi_{zt} = \dfrac{2a}{r^5} z
$$

$$
\Gamma^t_{z\varphi} = -\dfrac{2a^3}{r^5} z
$$

Since all the Christoffel symbols are proportional to $z$, one may rewrite the
expression for the vertical acceleration as:

\begin{equation}\label{E:gzo}
g_z = - \Omega_z^2(r,a) z
\end{equation}

where $\Omega_z(r,a)$ is the local co-moving frequency of vertical
oscillations. { Following \citet{ALP97}, one can substitute the above
expressions for Christoffel symbols and metric components and
express $\Omega_z$ in the
compact and elegant form independent of radial velocity:

\begin{equation}\label{E:gz}
\Omega_z^2 = \dfrac{1}{r^4}\left( u_\varphi^2 - a^2 \left(u_t^2 - 1\right)\right)
\end{equation}

\section{Acceleration caused by magnetic field tension}\label{sec:mag}

\subsection{Magnetic field components}

 Let us consider the case of ideal electrodynamics where the electric
  field is exactly zero in the comoving frame and thus $u_k F^{ik}=0$,
  where $F^{ik}$ is electromagnetic tensor. Electromagnetic fields in this
  case may be described solely by one co-moving magnetic field four-vector
  $B^i$, $B^i = u_k\ ^*\!F^{ik}$, where $ ^*\!F^{ik} = e^{ikjl} F_{jl}$ is
  dual electromagnetic tensor. Such description was used, for example, in
  \citet{krolik99, komi06}. The dual tensor itself may be expressed as
  $^*\!F^{ik} = u^i B^k - u^k B^i $. 
It may be shown that the scalar
product $u_i B^i =0$ meaning that the magnetic field has zero temporal
component in the co-moving frame. As long as $u_z \ll u_{r} \sim u_\varphi
\sim u_t$, one can
safely neglect the vertical components of the field when expressing the 
temporal component $B^t = -\dfrac{1}{u_t}\left( u_r B^r + u_\varphi B^\varphi
\right) $. 

The first pair of Maxwell equations may be written as (see for example
\citet{landafshitz}, chapter IV):

$$
(^*\!F^{ik})_{;k}=0,
$$

or, since the tensor is antisymmetric (ibid., chapter X):

$$
\left( \sqrt{-g}\times ^*\!\!F^{ik} \right)_{,k} =0,
$$

where $\sqrt{-g} \simeq r^2$
near the equator. Integrating over the solid angle ($\theta$ and $\varphi$
coordinates) and taking into account
the considered configuration should be axisymmetric and stationary yields the
following system of invariants:

\begin{equation}\label{E:Bgen}
4\pi H r  \left( u^i B^r - u^r B^i\right) =  \Phi^{r i},
\end{equation}

 It is assumed that dynamically important magnetic fields are concentrated
inside the accretion flow of thickness $H$. Since magnetic fields are advected
by accreting matter one should not expect to find strong magnetic fields far
from the equatorial plane. 
Radial magnetic field is easily obtained by contracting (\ref{E:Bgen}) with
4-velocity, bearing in mind that $u_t$ and $u_\varphi$ are constant and radial
motions near the edge of the disc are negligible. 

\begin{equation}\label{E:Br}
B^r(r) = \dfrac{H_0 r_0}{Hr}
B^r_0 
\end{equation}

 Let us first consider magnetic field that is purely poloidal in the disc.
Then,  substitution of $i=\varphi$ into (\ref{E:Bgen}) yields:

\begin{equation}\label{E:Bphi}
\begin{array}{l}
B^\varphi(r) =  \dfrac{H_0 r_0}{Hr u^r} \times \left[ u_0^r B_0^\varphi +
  \left(u^\varphi - u_0^\varphi\right) B_0^r\right] =\\ 
\qquad{} = \dfrac{H_0 r_0}{Hr } \times 
  \dfrac{u^\varphi - u_0^\varphi}{u^r} B_0^r\\
\end{array}
\end{equation}

Hence for the temporal component:

\begin{equation}\label{E:Bt}
\begin{array}{l}
B^t(r) =  -\dfrac{H_0 r_0}{Hr u^r u_t} \times \left[ 
\left( u_r u^r + u_\varphi (u^\varphi - u_0^\varphi) \right)B_0^r + u^r_0 u_\varphi B_0^\varphi
\right] =\\
\qquad{} =  -\dfrac{H_0 r_0}{Hr} \times \dfrac{u_r u^r + u_\varphi (u^\varphi
  - u_0^\varphi)}{u^r u_t} B_0^r
\end{array}
\end{equation}

Approximately, these equalities work well whenever we are far enough from the
disc edge. Note
that initial toroidal field enters the above expressions in combination $u^r_0
B^\varphi_0$ hence one may adopt $B^\varphi_0 \simeq 0$ as well as
$B^z_0\simeq 0$ since:

\begin{equation}\label{E:Bz}
B^z(r) =  \dfrac{H_0 r_0 u^r_0}{Hr u^r}\times B^z_0 
\end{equation}

It is natural to assume that
near the ISCO, magnetic field components are of the same order of magnitude,
$\sqrt{g_{\varphi\varphi}} B^\varphi \sim \sqrt{g_{rr}} B^r$, while the radial
velocity is much smaller than the azimuthal.
Further inside, where
$u^{\hat{\varphi}} \sim u^{\hat{r}} $, azimuthal field becomes much
larger and $B^\varphi \sim \dfrac{u^\varphi}{u^r} B^r$. Azimuthal field behaves
reasonable near the ISCO since the square bracket in (\ref{E:Bphi}) approaches
zero when $r\to r_{ISCO}$. This azimuthal field may be considered wound up by
differential rotation. 

It is however instructive to consider the opposite case when initial toroidal
fields dominate.  If initially radial magnetic field $B^r_0=0$, radial magnetic
fields are also absent at smaller radii ($B^r =0$), and azimuthal and temporal
components of initially purely toroidal field equal:

\begin{equation}\label{E:Bphi:tor}
B^\varphi_{tor} = \dfrac{H_0 r_0}{Hr} \dfrac{u_0^r}{u^r} B^\varphi_0
\end{equation}

\begin{equation}\label{E:Bt:tor}
B^t_{tor} = -\dfrac{H_0 r_0}{Hr} \dfrac{u_0^r}{u^r} \dfrac{u_\varphi}{u_t} B^\varphi_0
\end{equation}

In this case, the magnetic field strength is proportional to the initial
radial velocity and is thus important either in thicker discs or when toroidal
fields are much stronger than poloidal. Two-dimensional simulations
  suggest that magnetic fields in the disc do have such anisotropy, see
  section \ref{sec:harm:res}.

\subsubsection{Seed fields in the disc}

To estimate the initial magnetic field for the
important case of standard disc, one can adopt equipartition in the form
$p_{M0} = p_0 / \beta$, where $p_0$ is the characteristic value of standard
disc pressure near the ISCO:

\begin{equation}\label{E:pm0}
p_{M0} / \dot{M} \simeq \dfrac{3}{8\pi \alpha\beta} \dfrac{\Omega_K^0}{H_0}
\end{equation}

Assuming the field  isotropic, one can obtain an estimate for the radial field in
all the standard disc regimes:

\begin{equation}\label{E:b0}
B^{\hat{r}}_0 \simeq \sqrt{\dfrac{\Omega_K^0 \dot{M}}{\alpha\beta H_0}}
\end{equation}

This estimate will be used later for numerical calculations. Since the
acceleration created by magnetic fields $a_B \propto B^2/\dot{M}$ (see next
sub-section), this result
is quite universal and does not directly depend on $\dot{M}$. Physical
component of $B^z_0$ is expected to be of the same order, but since the impact
of the vertical fields is diminished by the flow rarefaction, in numerical
calculations I neglect vertical fields. 

\subsection{Vertical acceleration}

As it was shown in section \ref{sec:basic}, the motion of a test particle at
the surface of the flow is determined by vertical gravity and
magnetic stress.
Let us leave only two constituents in $T^{ik} \simeq \rho u^i u^k +
T_{EM}^{ik}$ and assume the electromagnetic part of the stress-energy tensor has
only one component $T_{EM}^{zz}$ important for the vertical structure of the
disc equal to:

$$
\begin{array}{l}
4\pi T_{EM\ z}^{z} = F^{zj}F_{jz} -\dfrac{1}{4}F^{ik} F_{ik}  = \\
\qquad{} = \dfrac{1}{2}\left(B^2 - 2 B_z^2\right) = \dfrac{1}{2}\left(B_r B^r + B_\varphi
B^\varphi + B_t B^t - B_z B^z\right) \\
\end{array}
$$

Or, using physical magnetic field components:

\begin{equation}\label{E:temzz}
-4\pi T_{EM\ z}^{z} = \dfrac{1}{2} \left( B_{\hat{\varphi}}^2 +
B_{\hat{r}}^2 + B_{\hat{t}}^2 - B_{\hat{z}}^2 \right)
\end{equation}

Vertical acceleration due to magnetic pressure { (vertical gradient here is
approximated as $d/dz \simeq 1/H$)}:

\begin{equation}\label{E:abz}
a_B^z = u^k u^z_{;k} \simeq \dfrac{1}{\rho} \dfrac{d}{dz}\left(
T_{EM\ z}^{z}\right) \simeq - \dfrac{4\pi T^z_z r u^r}{\dot{M}} 
\end{equation}

where the energy-stress tensor component is estimated by the above equation
(\ref{E:temzz}), and the magnetic field component values are calculated
following
the equations (\ref{E:Br}-\ref{E:Bt}) or (\ref{E:Bphi:tor}-\ref{E:Bt:tor}) of
the previous sub-section. 

\subsection{Equilibrium disc thickness}\label{sec:zeq}

\subsubsection{Magnetically supported disc}

The condition of vertical equilibrium may be written as:

$$
a_B^z = -g_z
$$

Above we have seen (equation (\ref{E:gzo})) that $g_z \propto z$. Magnetic
field acceleration, on the other hand, is proportional to compression
factor squared $\propto (H_0/H)^2$, with a coefficient depending on $r$ and
$a$ and initial fields. This may be written as:

$$
\Omega_z^2(r,a) H =  \dfrac{K(r,a)}{\alpha\beta} \dfrac{H_0}{H^{2}},
$$

where $K(r,a)$ is a certain function of radius and Kerr parameter. 

\begin{equation}\label{E:zeq}
H_{eq} = \left( \dfrac{H_0 K(r,a)}{\alpha \beta \Omega_z^2(r,a)} \right)^{1/3}
\end{equation}

Highly dissipative flow will have a disc half-thickness close to the
equilibrium vertical scale $H_{eq}$,
otherwise it will oscillate near this value (see below
section~\ref{sec:osc}). 

Scaling $H_{eq} \propto H_0^{1/3}$ suggests that for a disc thin enough,
$H_{eq}> H_0$ and the inner flow of a thin disc
should be comparably geometrically thick.
It should be also noted that $a_B^z \propto \beta^{-1}$ hence $H_{eq} \propto
\beta^{-1/3}$. { On the other hand, for relatively thick discs,} high $\alpha
\beta \gg 1$ are needed to prevent
the disc height exceeding $r$ hence I restrict myself to the very large
value of $\beta =100$ when solving the equations numerically.

\subsubsection{Thermal pressure}

If vertical gravity can not be balanced by magnetic fields, the free fall in
vertical direction is stopped by increasing pressure of the compressed
gas. For small enough rotation parameters and initial radial
velocities, the
radial free-fall times are long enough to approach the equilibrium state when:

$$
\Omega_z^2 H_{m} \simeq \dfrac{p_c}{\rho_c H_{m}},
$$

where $\rho_c$ and $p_c$ are density and pressure near the equatorial plane, and $H_m$ is
the vertical scale,

\begin{equation}\label{E:hm}
H_m \simeq \sqrt{\dfrac{p_c}{\rho_c}}\Omega_z^{-1}
\end{equation}

Pressure and density here are determined by compression and decompression
processes in the flow as well as by losses to radiation that should be
important in reality but are not included in the simulations presented below
(section \ref{sec:harm}) in this work. 


\section{Numerical solution}\label{sec:res}

\subsection{The Cauchy problem}\label{sec:cau}

Let us consider a free-falling inner disc with a density independent of
vertical coordinate in the range $-H < z
< H$. Such a disc would preserve its vertical structure, and its evolution
with time (and $r$) is self-similar in the sense that all the particle
trajectories differ only in some multiplication factor in vertical direction. 
{ Vertical four-velocity component $u^z$ is thus proportional to $z$, and
  it is sufficient to consider the motion of the particles consisting the
  surface of the inner disc ($z=H$). The thickness itself is $H=\int u^z d\tau$.} 
Thickness as a function of radius obeys the equation:

\begin{equation}\label{E:hevol}
u^r\dfrac{d}{dr}\left( u^r \dfrac{dH}{dr} \right) = g_z+a_B
\end{equation}

where the accelerations $g_z$ and $a_B$ are calculated according to equations
(\ref{E:gzo}) and (\ref{E:abz}), respectively. 
Solution is obtained by solving the Cauchy
problem for this equation where all the coefficients depend on $r$ and $H$.
The initial conditions are $H(r_{ISCO}) = H_0$ and $dH/dr(r_{ISCO}) =0$.

For the two right-hand terms in (\ref{E:hevol}), scalings with disc height are
different: $g_z
\propto H$ and $a_B \propto H^{-2}$. At each radius value, there is an
equilibrium thickness (see section~\ref{sec:zeq}), 
and the solution of the above equation in general case
is oscillating around the equilibrium value. Oscillation frequency is set
mainly by the gravity term and is thus close to $\Omega_z(r,a)$. 
Velocity components $u_\varphi$ and $u_t$ are assumed
constant and equal to the values at the last stable orbit, and $u^r$ is
estimated using the normalization condition ($u_i u^i =-1$) modified to make
the evolution near the disc boundary smoother:

\begin{equation}\label{E:urreal}
u_r(r) \simeq \sqrt{-g_{rr}(g^{tt}u_t^2+2g^{\varphi t}u_tu_\varphi +
  g^{\varphi\varphi}u_\varphi^2+1)+u_{\hat{r},0}^2},
\end{equation}

where $u_{\hat{r},0}$ is the physical radial velocity at the ISCO that was
estimated (unless otherwise stated) 
according to the non-relativistic thin disc model as:

$$
u_{\hat{r},0} \simeq v_r \simeq \alpha \dfrac{H_0^2}{r_0} \Omega_K^0
$$

Far enough from ISCO, the flow is independent of the exact value of
initial radial velocity, especially if it is small. At ISCO, corrections to
energy and
momentum due to non-zero radial velocity are quadratic in $u_{\hat{r},0}$ and
may be generally ignored. 

\begin{figure*}
 \centering
\includegraphics[width=0.8\textwidth]{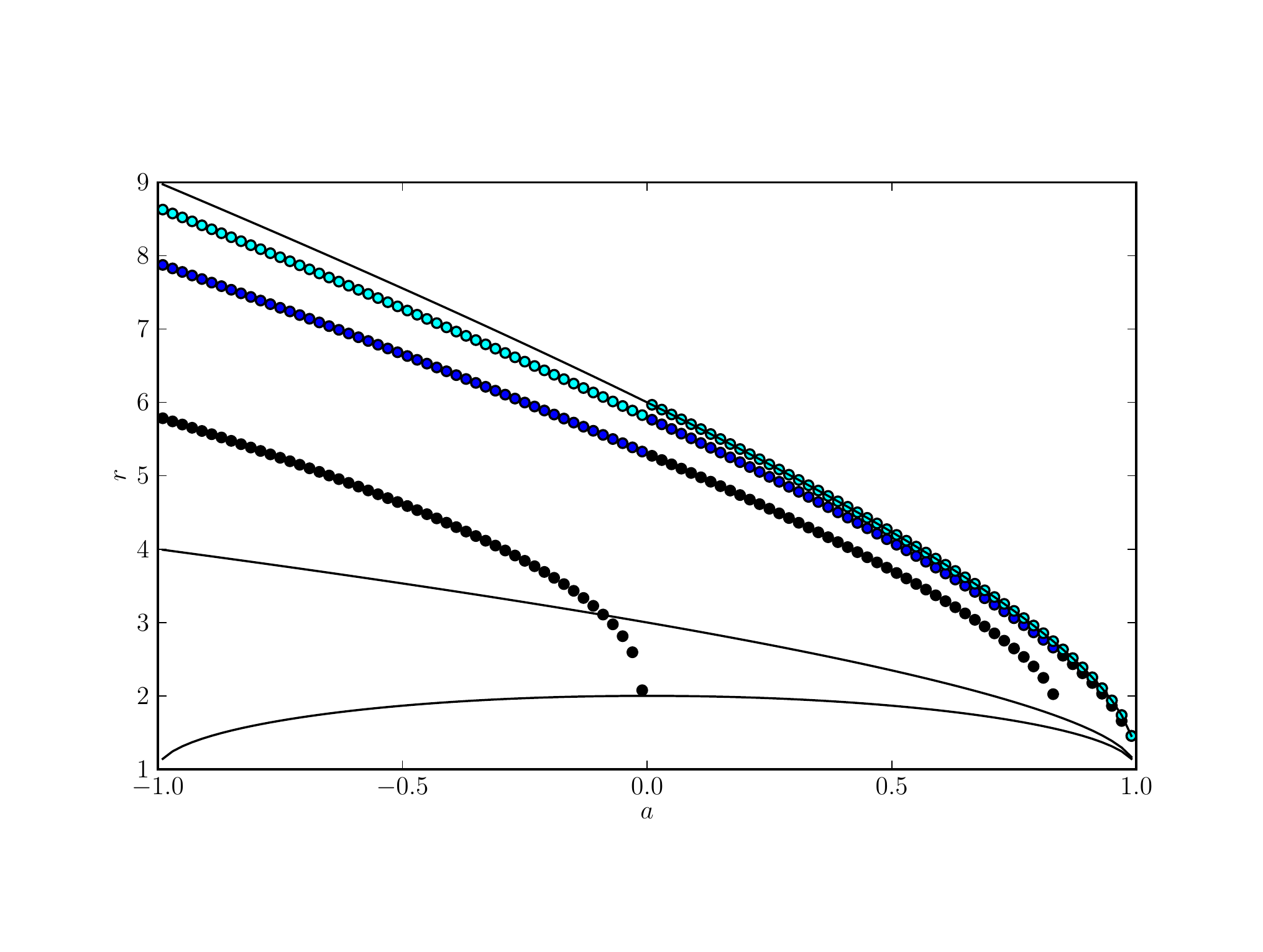}
\caption{ Innermost three nodes of ballistic (free-falling) solutions as
  functions of Kerr parameter. Characteristic radii are shown with solid
  curves: horizon $r_{hor}$ (the smallest), photon circular orbit radius
  $r_{ph}$ and last stable orbit radius $r_{ISCO}$ (the largest). } 
\label{fig:roots}
\end{figure*}

\subsection{Ballistic trajectories}\label{sec:ff}

If no pressure sources are present and the trajectories are allowed to
intersect each other, one arrives to the case considered by
\citet{betch05}. The flow intersects itself at certain distances, and the
number of nodes decreases with Kerr parameter and initial radial velocity. 
Positions of the three innermost nodes are shown in figure
\ref{fig:roots} for $u_{r,0}=0$. 

Adding small magnetic field or other pressure source transforms the nodes into
shocks with Mach numbers essentially close to unity because the vertical
motions are expected to stop when $p \simeq \rho v_z^2$ where $p$ includes
both magnetic and thermal pressure in vertical direction.

\subsection{Equilibrium disc height}\label{sec:res:zeq}

\begin{figure*}
 \centering
\includegraphics[width=0.9\textwidth]{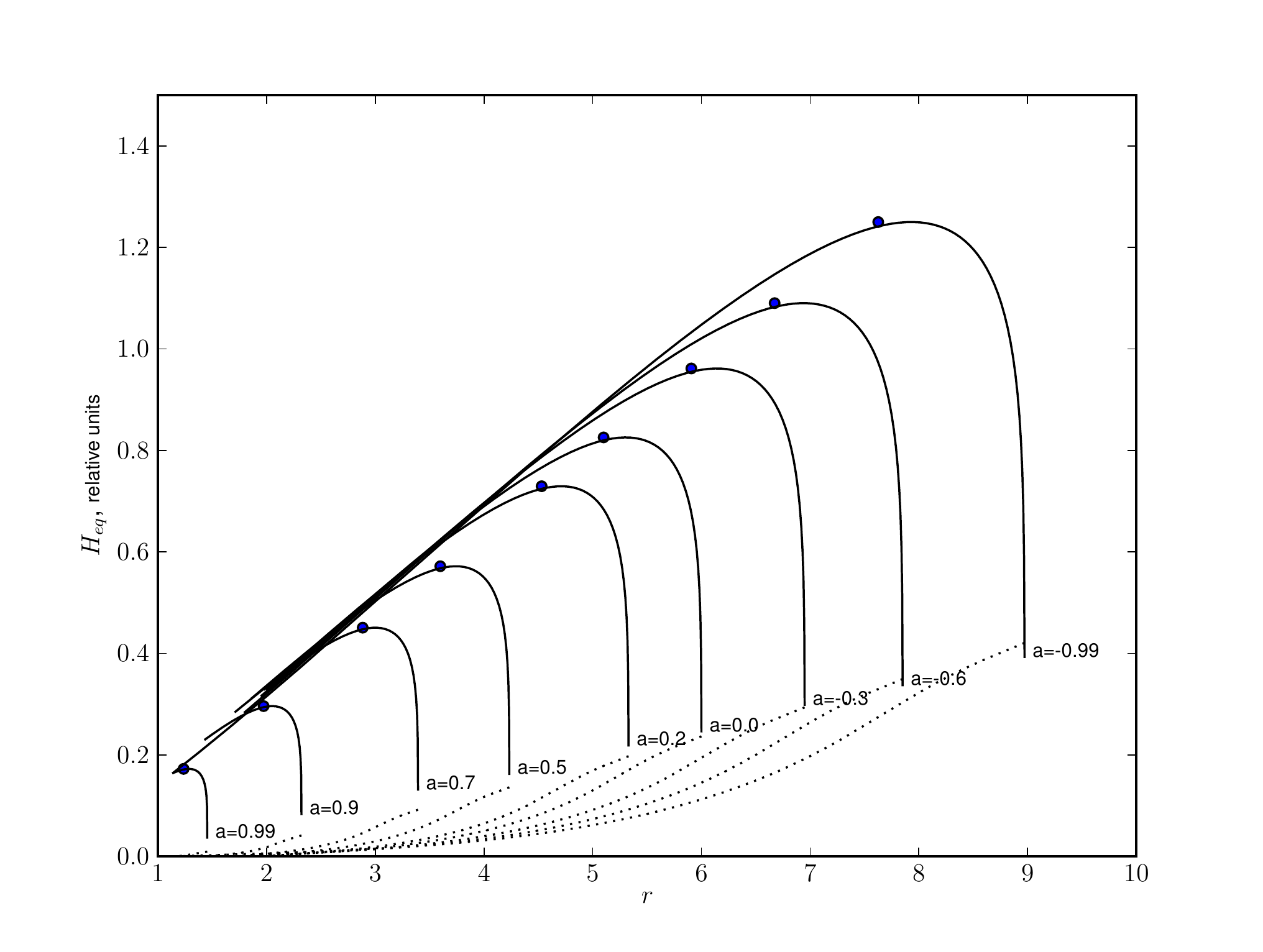}
\caption{ Equilibrium disc thicknesses for different values of Kerr parameter.
In this graph, $\alpha \beta=100$ and $h_0 = 0.1$ is assumed. Solid and dotted
lines correspond to radial and toroidal seed magnetic fields. } 
\label{fig:zeqs}
\end{figure*}

\begin{figure*}
 \centering
\includegraphics[width=1.1\textwidth]{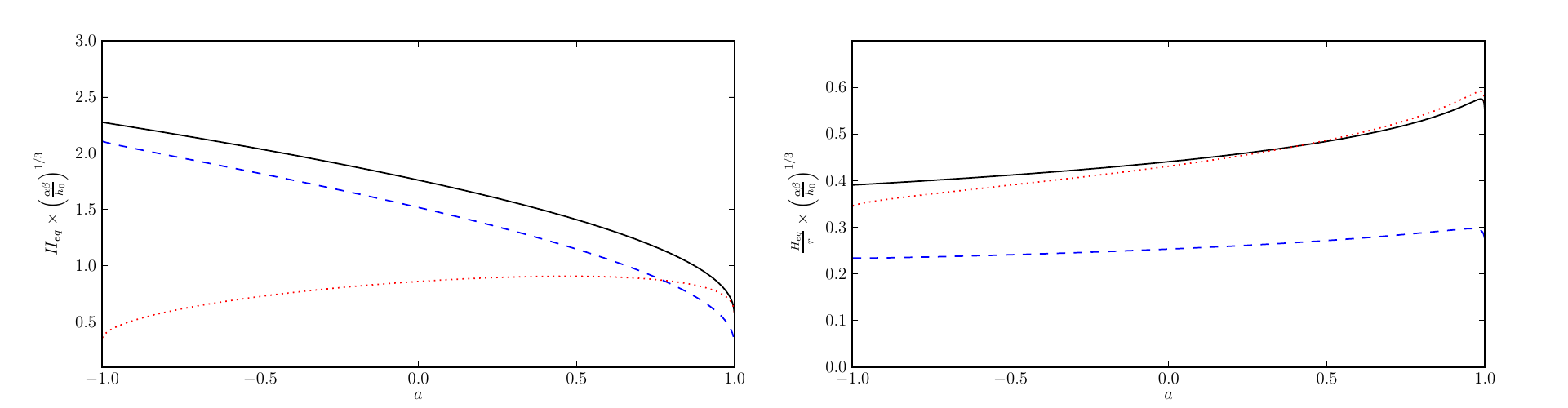}
\caption{
Normalized equilibrium disc thicknesses as a function of Kerr
  parameter at
  three characteristic radii: midway between $r_{ISCO}$ and horizon (solid
  line), near $r_{ISCO}$ (blue/dashed) and near $r_{hor}$ (red/dotted).
  Again, $\alpha \beta=100$ and $h_0 = 0.1$ is assumed. The left panel shows
  absolute, and the right relative thicknesses. Purely radial seed fields were
  considered.} 
\label{fig:zeqmesh}
\end{figure*}

\begin{figure*}
 \centering
\includegraphics[width=1.1\textwidth]{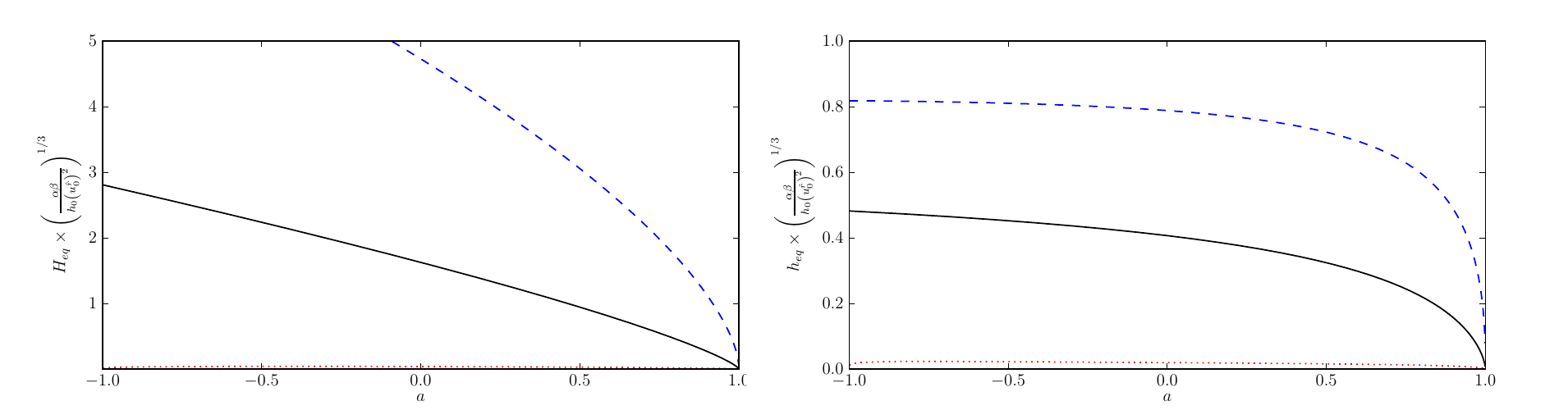}
\caption{Same as previous figure, but initially 
  toroidal seed fields were considered
  instead of radial. Dotted curve was calculated for the radial coordinate
  equal to $1.05r_{hor}$. } 
\label{fig:torzeqmesh}
\end{figure*}

As it was shown in section \ref{sec:zeq}, equilibrium disc thickness
scales $\propto \left( h_0 / \alpha \beta \right)^{1/3}$, where $h_0 =
H_0/r_0$ is relative initial half-thickness of the disc. The equilibrium
quantity $H_{eq}$ is a
function of radius with a maximum at $R \sim 0.85 r_{ISCO}$, weakly dependent on
Kerr parameter (see figure \ref{fig:zeqs}). Equilibrium thicknesses at
different radii as functions of Kerr parameter are shown in figures
\ref{fig:zeqmesh} and \ref{fig:torzeqmesh}, correspondingly. 
Relative disc thickness $h_{eq}=H_{eq}/r$ is a much weaker function of $a$. 

If seed magnetic fields are primarily poloidal, numerical
results allow to make the following estimate for the equilibrium disc
thickness near $r_{ISCO}$, slightly dependent on $a$:

\begin{equation}\label{E:eq:polout}
h_{eq,\,out,\,pol} \simeq \left( 0.2 \div 0.3 \right)
\left( \dfrac{h_0}{\alpha\beta}\right)^{1/3}
\end{equation}

Near the horizon, a similar estimate may be made:

\begin{equation}\label{E:eq:polin}
h_{eq,\,in,\,pol} \simeq \left( 0.35 \div 0.6 \right)
\left( \dfrac{h_0}{\alpha\beta}\right)^{1/3}
\end{equation}

Equilibrium relative disc thickness at intermediate radii is much closer to
the estimate at the horizon, as one can see in figure~\ref{fig:zeqmesh}. 

If toroidal fields dominate in the accretion disc and thus determine the
structure of the flow inside the ISCO, the equilibrium thickness approaches
zero when $a\to 1$ and $r\to r_{hor}$ (see figure~\ref{fig:zeqs}). For
relatively low $a \lesssim 0.7$ and for counter-rotation, initial thickness of
the flow supported by toroidal fields is:

\begin{equation}\label{E:eq:torout}
h_{eq,\,out,\, tor} \simeq \left( 0.7 \div 0.8 \right)
\left( \dfrac{h_0 \left(u_0^{\hat{r}}\right)^2}{\alpha\beta}\right)^{1/3}
\end{equation}

If the initial toroidal fields are large, this formula is a reasonable
estimate for the initial thickness, but to support the inner parts of the flow
one needs to consider seed radial fields. 

\subsection{Height oscillations}\label{sec:osc}

Clearly, equation~(\ref{E:hevol}) describes disc thickness oscillations and
has one neutral focus near $H=H_{eq}$. If vertical motions are subject to any
kind of dissipation (shock waves, bulk viscosity etc.), the oscillations are
damped.
In figure \ref{fig:osc}, disc thickness oscillations with radius are shown
for different Kerr
parameters. Self-intersecting dust free-fall solutions are shown with dots,
and dashes mark the equilibrium disc thickness. Rapid oscillations near the
ISCO occur only for smaller initial radial velocities. 

\begin{figure*}
 \centering
\includegraphics[width=0.75\textwidth]{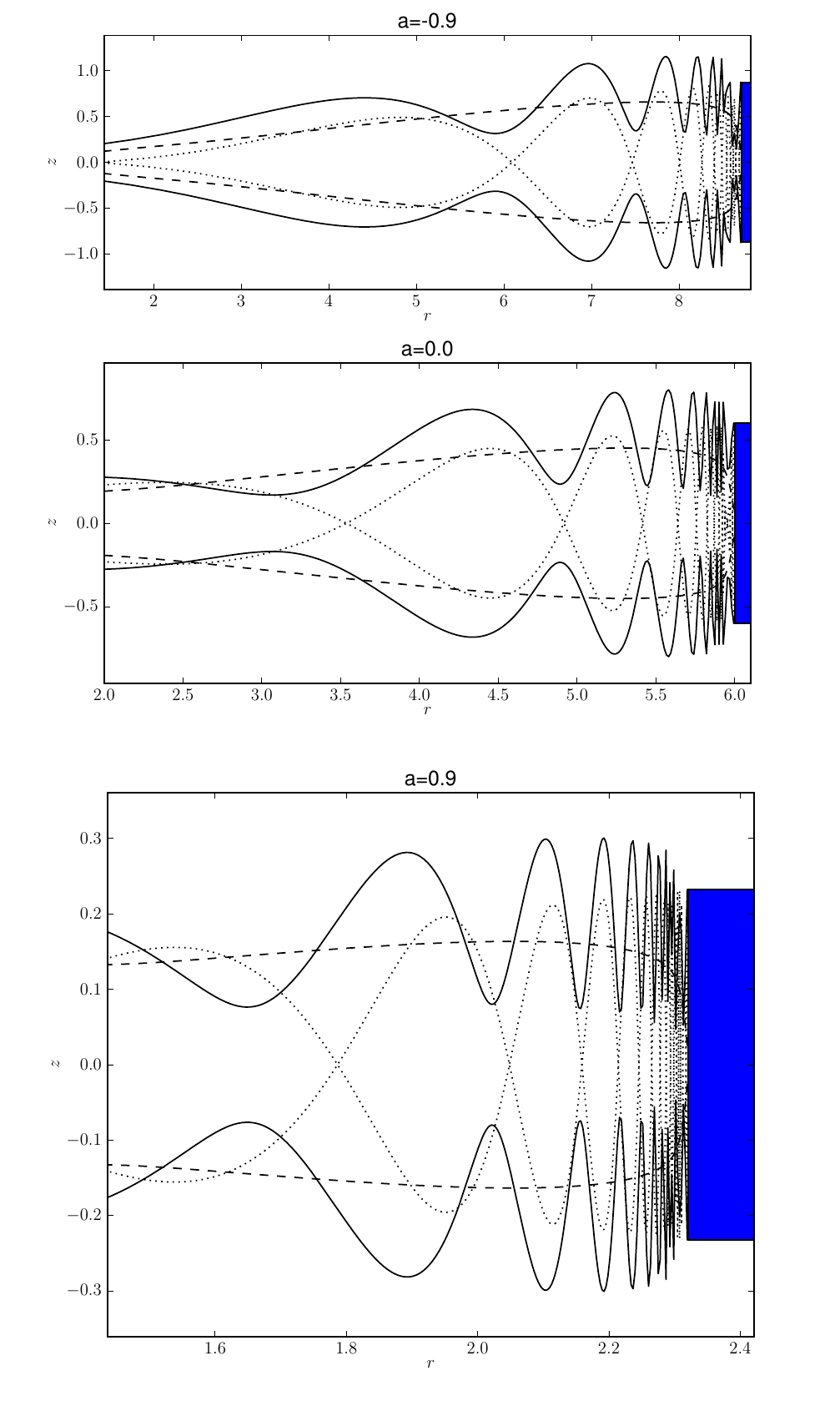}
\caption{ Disc thickness oscillations. Dotted lines show test particle
  trajectories  in absence of pressure
  (with self-intersections), dashed lines mark the equilibrium disc thickness,
  and solid lines show magnetically supported disc thickness dynamical
  evolution. The bars on the right show schematically the initial thickness of
  the disc. { Initial disc thickness everywhere is set to $h_0=0.1$.} } 
\label{fig:osc}
\end{figure*}

\section{Numerical simulations in 2D}\label{sec:harm}

\subsection{Basic information and code setup}

{\it HARM} is a publicly available code that allows to consider MHD
problems in full General Relativity in a fixed Kerr background
metric \citep{harm1,harm2}. 
It was successfully applied to problems like Blandford-Znajek
magnetosphere structure and jet formation, magneto-rotational
instability development in accretion discs and tori and some others. 
It does not include implicitly any
dissipation mechanisms such as magnetic reconnections and fluid viscosity but
they effectively emerge at small spatial scales due to numerical
viscosity. Radiation is neither included in the code hence a viscous disc
conserves its thickness and does not evolve toward the thin disc solution. 

Currently, there is fair
understanding of what happens to a regular vertical magnetic field frozen into a thick
disc (see for example \citet{harm3}). 
Black hole then becomes immersed in a
regular magnetic field with the strength determined by flux accumulation. 
If magnetic field
instead is chaotic and its spatial scales are small enough,
the strength of regular fields through the black hole horizon are smaller than
that of the fields in the inner disc and accretion flow.  The critical
correlation length may be estimated by comparing the regular and chaotic
magnetic field components near the horizon, $B^{reg}$ and $B^{ch}$, where:

$$
B^{reg} \simeq \frac{\Phi}{\pi r_0^2} \simeq B_0 \times \dfrac{L_{reg}^2}{r_0^2},
$$

where $L_{reg}$ is characteristic magnetic field correlation scale in the
disc, and:

$$
B^{ch} \simeq \dfrac{H_0 r_0}{H_{in} r_{in}} B_0
$$

Regular field dominates if the fields in the inner parts of the accretion
disc are correlated on spatial scales about of larger than:

\begin{equation}\label{E:lreg}
L_{reg} \sim \sqrt{\dfrac{H_0 r^3_0}{H_{in} r_{in}}} \gtrsim r_0
\end{equation}

Here, the ``zero'' index corresponds to the last stable orbit, and 
``in'' to the quantities near the horizon. 

To check the viability of the analytical approach to the inner flow, I
formulate a series of two-dimensional {\it HARM} problems with the following initial conditions.
Initially, an aligned torus orbiting a rotating black hole was placed between
the last stable orbit and approximately 50$GM/c^2$. Exact solution of
\cite{FMtorus} was chosen, one of the standard and well-tested forms of
initial matter distribution. Torus size parameters were chosen depending on
the Kerr parameter (see table \ref{tab:harm:mod}), and $\varkappa$ parameter
was everywhere set to 1. 

Poloidal initial magnetic fields were set through vector potential having
only one (azimuthal) non-zero component $A_\varphi$:

$$
\vector{B}_{pol} = \nabla \times \vector{A} 
$$

$$
B^r = - \dfrac{1}{\sqrt{-g}} \pardirone{\theta}{A_\varphi}
$$

$$
B^\theta = - \dfrac{1}{\sqrt{-g}} \pardirone{r}{ A_\varphi},
$$

where $g=-\rho^4\sin^2\theta = -(r^2+a^2\cos^2\theta)^2\sin^2\theta$ is metric determinant.

\subsubsection{Single loop case}

The above estimate (\ref{E:lreg}) means that even smooth
seed magnetic fields
with spatial gradients $\sim B/r$ produce stronger magnetic fields in the
accretion flow than the regular vertical field component.
Thus the first case I
consider in four of the six simulation runs 
is a single-loop scenario with $A_\varphi$ 
set proportional to the local pressure in the
torus. This configuration is also standard, it was used for example in
\citet{shafee}. 

Three models A1, B1 and C1 reproduce accretion upon black holes with different
Kerr parameters (see table~\ref{tab:harm:mod} for details). 
Counter-rotation is described via $a<0$. Dependence on
resolution was considered by performing one additional simulation, B1h. 

\subsubsection{Multi-loop case}

On the other hand, magnetic fields may be small-scale and chaotic. Stochastic
poloidal field configuration may be thought of as produced by a number of
toroidal currents randomly placed in the disc. 
 Vector potential produced by such a current in the non-relativistic case is given by the following
expression:

$$
A_\varphi^{loop} \propto \sqrt{\dfrac{r_{loop}}{r}} \times \left(
\left(\dfrac{2}{k}-k\right) {\rm K}(k) - \dfrac{2}{k} {\rm E}(k) \right),
$$

where 

$$
k^2= \dfrac{4r_{loop}r}{(r+r_{loop})^2+(z-z_{loop})^2}
$$ 

and ${\rm E}(k)$ and ${\rm K}(k)$ are complete elliptic integrals of the first
and the second kind
(this solution is given, for example, in \citet{BT78}). Though the solution
for a single current loop is not exact in general relativity, 
it corresponds to some current configuration not too much different from a
single loop. The worst disadvantage of this solution is that it creates
magnetic stresses that are far from proportionality with the gas energy and
momentum density. To soften this effect, infinitely-thin loops were replaced
by flux ``tubes'' with vector potential multiplied by the factor:

$$
A_\varphi^{tube} = A_\varphi^{loop} \times \dfrac{d}{\sqrt{1+d^4}},
$$

where $d=\sqrt{(r-r_{loop})^2 + (z-z_{loop})^2}$. 

Poloidal magnetic fields were set as a
sum of 50 toroidal current tubes with random current scaling with the local
gas pressure and randomly placed inside the torus. The two simulation runs
differ in symmetry: in one case (A50s), vector potential was made symmetrical with
respect to the orbital plane, in the other case
anti-symmetrical. 

\bigskip

In both scenarios, the fields were scaled proportional to the 
local gas pressure with the initial
$\beta = 2000$. I set no initial toroidal fields since poloidal fields are
rapidly (on the time scales of several $GM/c^2$) wound up and converted to
toroidal. 

In total, 6 simulation runs were performed with different Kerr parameter
values (see table~\ref{tab:harm:mod}). Except for one run, the grid was chosen with 128 points in
radial direction and 192 in polar angle that provides $\Delta \theta \sim
10^{-2}$ resolution in polar angle. One of the simulations, B1, was repeated at
higher resolution (B1h has resolution 192$\times$320 instead of 128$\times$192).

The initial poloidal $\beta$ value was set to
2000, but the magnetic fields are substantially amplified by differential
rotation, turbulence and instabilities. 
For every run, there is an initial
period when only very weakly magnetized matter is accreted, and intense
accretion of the magnetized matter from the disc 
starts after several hundreds of $GM/c^3$. { All the quantities
given in tables \ref{tab:harm:res} and \ref{tab:harm:htor} were 
averaged over the time ranges (given in table \ref{tab:harm:mod} for every
model) when accretion was relatively stable.} 

\subsection{Results and comparison to analytical estimates}\label{sec:harm:res}

\subsubsection{Overview}

Main simulation results are given in tables
\ref{tab:harm:res} and \ref{tab:harm:htor}. Animations based on the results are available on
    {\it YouTube}, and the raw data may be provided upon request. 

Evolution of the magnetized torus follows more or less a single scenario:
first, the parts of magnetic loops extending beyond the torus rapidly expand
and straighten because in these parts the plasma is nearly
force-free. Inside the disc, magnetic fields are frozen-in and evolve slowly,
mainly due to differential rotation and magneto-rotational instability (MRI, see
\citet{MRI}). After thousands of dynamical times, 
magnetic field
amplification saturates, and quasi-steady accretion establishes. 
 The saturated magnetic fields are strongly anisotropic with
$B^{\hat{\varphi}} \gg B^{\hat{r}}$ but this may be an artifact of 2D
approach. Three-dimensional numerical simulations 
such as \citet{fromang07,parkin13} suggest that
saturated chaotic magnetic fields in accretion discs with Keplerian rotation 
are indeed anisotropic but with the toroidal fields
about one-two orders of magnitude stronger, not two-three orders as
two-dimensional simulations predict. 
{ Such a discrepancy is not unexpected. As the large-scale vertical 
magnetic field loses its energy due to magneto-rotational
  instability, the amplitudes of the radial and toroidal fields are determined
  by the secondary (parasitic) instabilities \citep{GX94} of the field
  configuration formed by the MRI as well as by differential rotation of the
  flow, Parker instability and reconnections \citep{TP92}. Parker instability
  in axisymmetric approximation is unable to affect the strength of toroidal
  magnetic field, while in 3D it may easily bend magnetic field lines in
  vertical direction and thus transfer azimuthal magnetic fields into
  vertical. 
}

For most of the simulations, toroidal
fields are strong enough to determine the inner flow structure. In
table~\ref{tab:harm:htor}, the larger value of the two equilibrium thicknesses
(determined by initially toroidal or initially radial fields) is
relevant as an estimate for the expected thickness of a magnetically supported
disc. 

In the
multiple-loop case, complicated magnetic field structure somehow delays
rapid field amplification and quasi-stationary accretion establishes later
(about $10^4$ from the start of the simulation). The start of quasi-stationary
accretion regime depends strongly on the starting magnetic field configuration
and varies from one run to another by a factor of several.

Viscosity parameter $\alpha$ was calculated as:

\begin{equation}\label{E:alpha}
\alpha = \left\langle \dfrac{T_{EM}^{\hat{r}\hat{\varphi}}}{p}\right\rangle = \left\langle  \sqrt{\dfrac{g_{rr}}{g_{\varphi\varphi}}}
\dfrac{T_{EM}^r\,_\varphi}{p}\right\rangle
\end{equation}

Averaging was performed near the equatorial plane in the inner part of the accretion
torus between $r_{ISCO}$ and $2r_{ISCO}$ (see $\alpha_{out}$ values in table
\ref{tab:harm:res}) and in the supersonic flow between the last stable orbit
and event horizon ($\alpha_{in}$). Viscosity parameter seems to be strongly
dependent on coordinates, being about unity inside the ISCO and above the disc
surface. I estimated the mean $\alpha$ acting in the flow by averaging the
above expression over
two disc heights inside the relevant range of radii.  Disc height itself
  was measured according to the technique described in section~\ref{sec:res:htor}. 

The resultant values of viscosity parameter in the disc are about several
percent or less.
However, further from the equatorial plane viscous stresses are about the same, while
thermal energy density decreases, and $\alpha$ may be, say, an order of
magnitude larger if larger distances (about $20^\circ$) from the equatorial plane are considered. The viscosity parameter behaviour near ISCO was considered with
the help of {\it HARM} in the work of \citet{shafee}. Their results are
qualitatively similar to mine. 

{
For different models, the product $\alpha \beta = \left\langle
\dfrac{2B^{\hat{r}}B^{\varphi}}{B^2}\right\rangle$ varies in the range
0.01$\div$0.2. Because azimuthal fields are much stronger, $\alpha \beta
\sim \sqrt{\dfrac{\beta_r}{\beta_\varphi}}$ (the third, vertical magnetic
field component is even smaller than the radial). 
Interpreted in terms of the ``tilt angle'' $\theta_B =
\dfrac{1}{2} \arcsin (\alpha \beta)$, the estimated values yield $\theta_B \sim
0.2\div 5^\circ$. Since comprehensive three-dimensional MHD simulations argue
for a universal value of $\theta_B \simeq 12\div 13^\circ$ \citep{sorathia},
we conclude that in real discs the contribution of azimuthal fields should be
smaller by a factor of several. 
}

\begin{table*}\centering
\caption{{\it HARM} model parameters. The time ranges (in $GM/c^3$) used for
  averaging are also given. Parameters of the initial Fishbone-Moncrief torus,
$r_{max}$ (radius of maximal density) and $r_{in}$ (inner radius) are also
  given. }
\label{tab:harm:mod}
\bigskip
\begin{tabular}{l|cccccc}
\medskip

Model & $a$  & $\beta_{init}$ & $r_{max}$ & $r_{in}$ &  number of loops & time
range \\

A1  & 0      & 2000 & 15.0   & 9.5  &  1  & 2000-5000 \\ 
B1  & 0.9     & 2000 & 8.4 & 4.8   &  1  & 3000-5700 \\ 
B1h (192$\times$320) & 0.9       & 2000 & 8.4   & 4.8  &  1 & 2000-5700\\ 
C1  & -0.9   & 2000 & 19.7 & 12.8  &  1  & 2500-4500 \\ 
A50s & 0      & 2000 & 19.7   & 12.8  &  50 & 10000-15000 \\
A50a & 0     & 2000 & 19.7   & 12.8  &  50 & 8000-10000 \\
\end{tabular}
\end{table*}

\begin{table*}\centering
\caption{{\it HARM} simulation results. The values of $\alpha$ and $\beta$ in
  the disc and inside the ISCO are given. Uncertainties are in fact root mean
  square deviations and reflect primarily the temporal variability of the
  quantities. }
\label{tab:harm:res}
\bigskip
\begin{tabular}{l|cccc}
\medskip
Model &  $\alpha_{out}$ & $\alpha_{in}$  &  $\beta_r$  &  $\beta_\varphi$ \\

A1  & 0.035$\pm$0.016 & 0.5$\pm$0.5  &  3000$\pm$3000  &  2.6$\pm$0.6  \\ 

B1  & 0.0085$\pm$0.0010 & 0.034$\pm$0.007  &  24000$\pm$10000  &  5.1$\pm$0.6   \\ 

B1h  & 0.010$\pm$0.003 & 0.020$\pm$0.014  &  $(1\pm 0.8)\times 10^4$   &  3.1$\pm$0.8  \\ 

C1  &  0.047$\pm$0.005 & 0.5$\pm$0.2  &  2000$\pm$300  &  3.9$\pm$1.2 \\ 

A50s  & 0.012$\pm$0.002 & 0.16$\pm$0.07  & $(5\pm 2)\times 10^4$ & 1.09$\pm$0.09 \\ 

A50a  & 0.00165$\pm$0.0002 & 0.04$\pm$0.05  &  $(2.0\pm 0.4)\times 10^5$ & 5.3$\pm$0.3  \\ 

\end{tabular}
\end{table*}

\begin{table*}\centering
\caption{Disc thicknesses estimated in the {\it HARM} simulations. Uncertainties are in fact root mean
  square deviations and reflect primarily the temporal variability of the
  quantities. Disc thicknesses given in this table are calculated for the two
  intervals, one outside and one inside ISCO, and  compared to the
  magnetically supported relative disc thickness near horizon and to the
  pressure supported disc thickness estimated for $r_{mb}$.}
\label{tab:harm:htor}
\bigskip
\begin{tabular}{l|ccccc}
\medskip
Model & $(1-2)r_{ISCO}$  & $r_{hor}-r_{mb}$  & \multicolumn{2}{c}{$\dfrac{H_{eq}}{r}(r_{hor})$}  & $\dfrac{H_{m}}{r}(r_{mb})$ \\
      &                 &                &  radial  &  toroidal &   \\

A1   & 0.064$\pm$0.008  &  0.04$\pm$0.04 &  0.078$\pm$0.011 & 0.20$\pm$0.03 & 0.034$\pm$0.002   \\ 

B1   & 0.108$\pm$0.005  &  0.025$\pm$0.007 &  0.063$\pm$0.004 & 0.025$\pm$0.003 & 0.049$\pm$0.002   \\ 

B1h  & 0.118$\pm$0.012  &  0.040$\pm$0.011 &  0.089$\pm$0.0014 & 0.029$\pm$0.003 &  0.047$\pm$0.004 \\ 

C1   & 0.090$\pm$0.003  &  0.030$\pm$0.13 & 0.073$\pm$0.004   & 0.36$\pm$0.03
& 0.0525$\pm$0.0015   \\ 

A50s & 0.086$\pm$0.004  &  0.023$\pm$0.008 &  0.040$\pm$0.004 & 0.41$\pm$0.01 &   0.0324$\pm$0.003    \\ 

A50a & 0.082$\pm$0.006  &  0.040$\pm$0.011  &  0.050$\pm$0.003 & 0.43$\pm$0.03 & 0.0283$\pm$0.0003   \\ 

\end{tabular}
\end{table*}

\begin{figure*}
 \centering
\includegraphics[height=0.9\textheight]{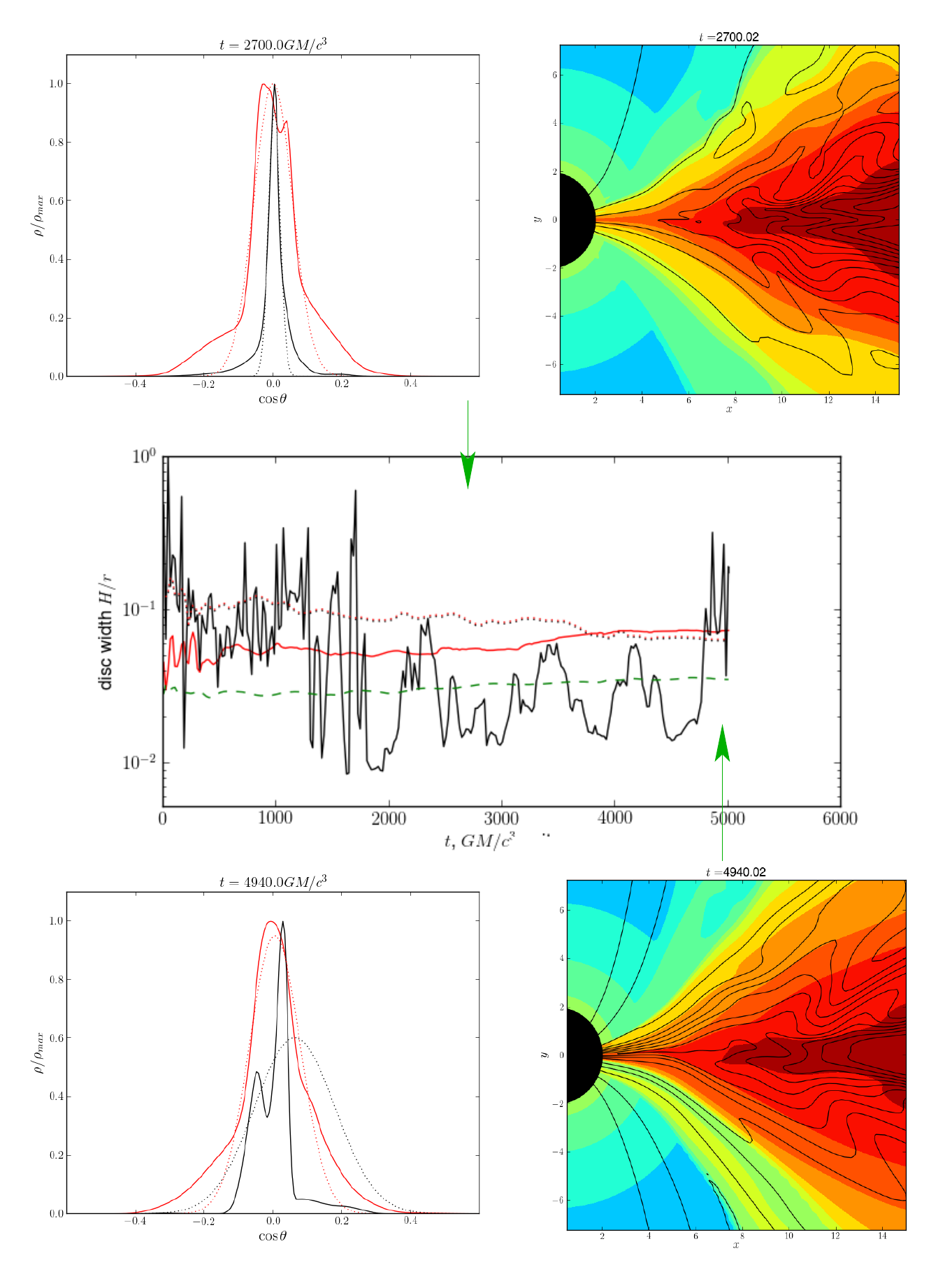}
\caption{ {\it Central:} disc relative 
  thicknesses for $a=0$ (A1 model) calculated 
  between 6 and 12 ~$GM/c^2$ (red/grey solid line)
  and between 2 and 6  ~$GM/c^2$ (black solid line). Equilibrium
  magnetically-supported disc
  thicknesses calculated for $r=r_{hor}$ (red/grey dotted line) and for
  $r=r_{ISCO}$ (black dotted) are also shown. { Green dashed curve shows the
  estimated equilibrium thickness of a disc supported by thermal pressure (at
  $r=r_{mb}$). } Upper and lower
  {\it right panels} show two representative snapshots at the instances shown by the
  arrows. Density is shown by colours/shades (logarithmic scale), and magnetic
  field lines are marked by contours.
 Corresponding radially-integrated density profiles are shown in the
  corresponding left panels: inner region ($r_{hor}..r_{mb}$) is shown with a
  black solid curve, disc region ($r_{hor}..r_{mb}$) with a red/grey solid
  curve. Dotted curves are Gaussian approximations. 
 } 
\label{fig:harm0}
\end{figure*}

\begin{figure*}
 \centering
\includegraphics[height=0.9\textheight]{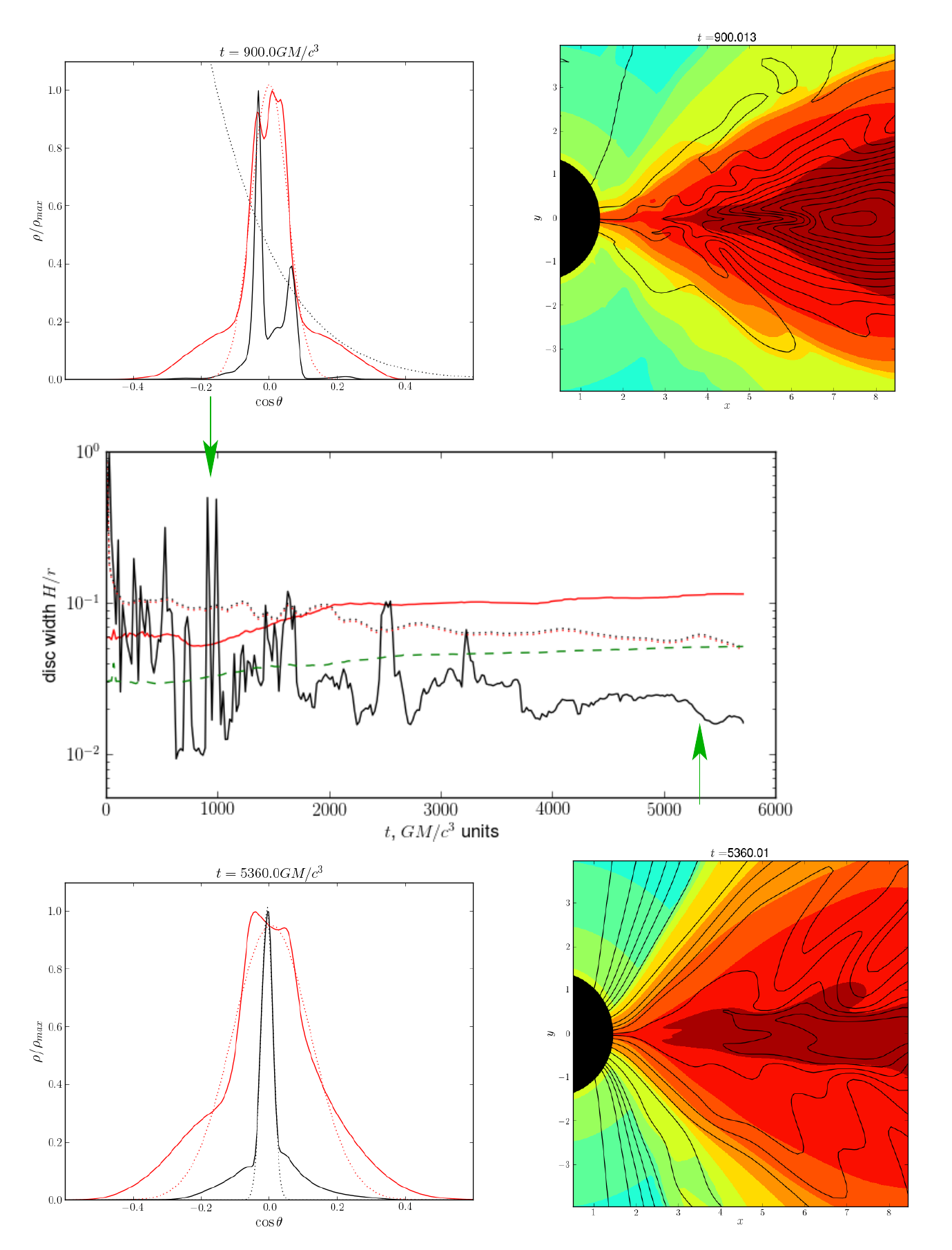}
\caption{Same as previous figure, but for $a=0.9$ (B1 model). } 
\label{fig:harm9}
\end{figure*}

\begin{figure*}
 \centering
\includegraphics[height=0.9\textheight]{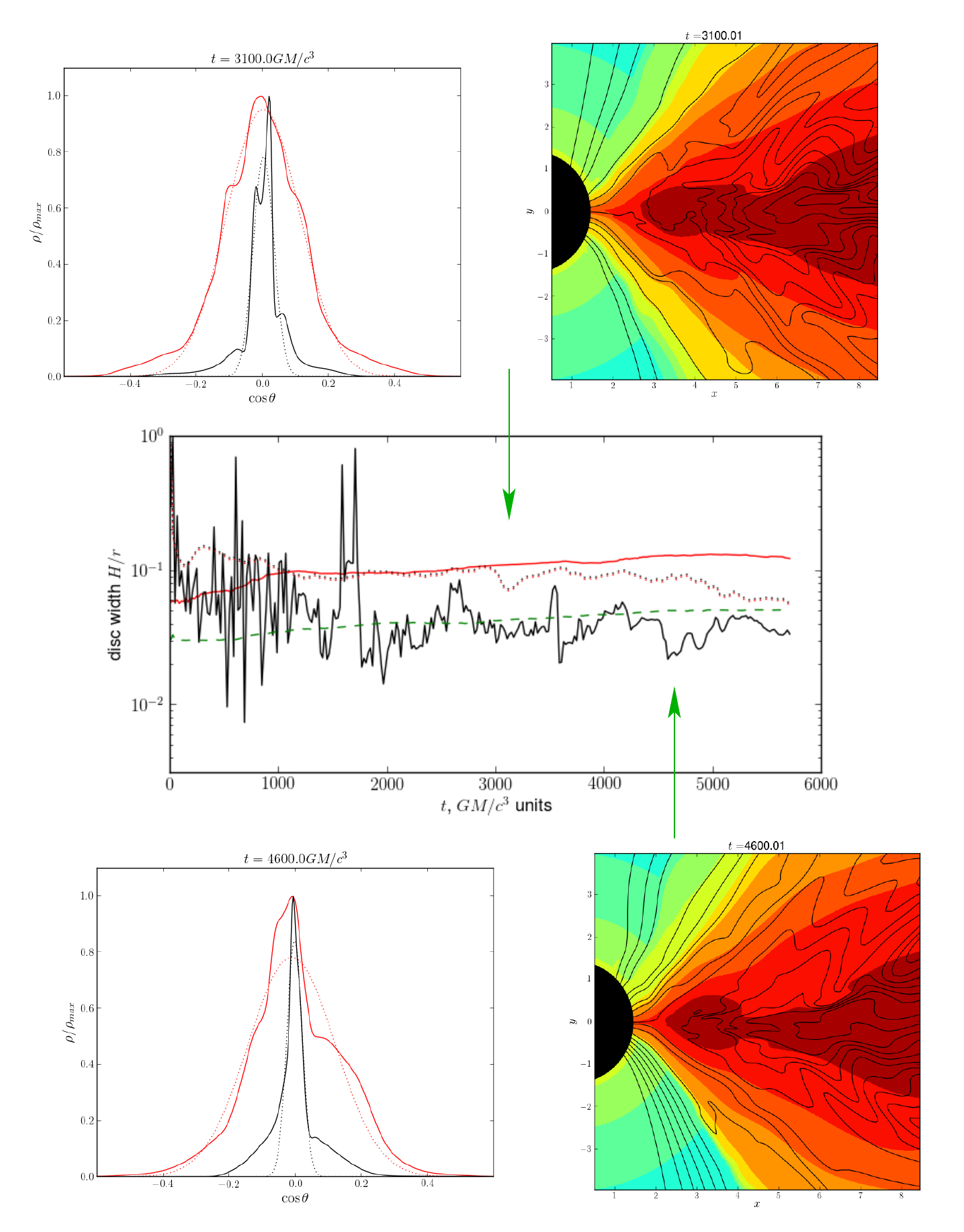}
\caption{Same as previous figure, but with higher resolution (B1h model, $a=0.9$). } 
\label{fig:harm9}
\end{figure*}

\begin{figure*}
 \centering
\includegraphics[height=0.9\textheight]{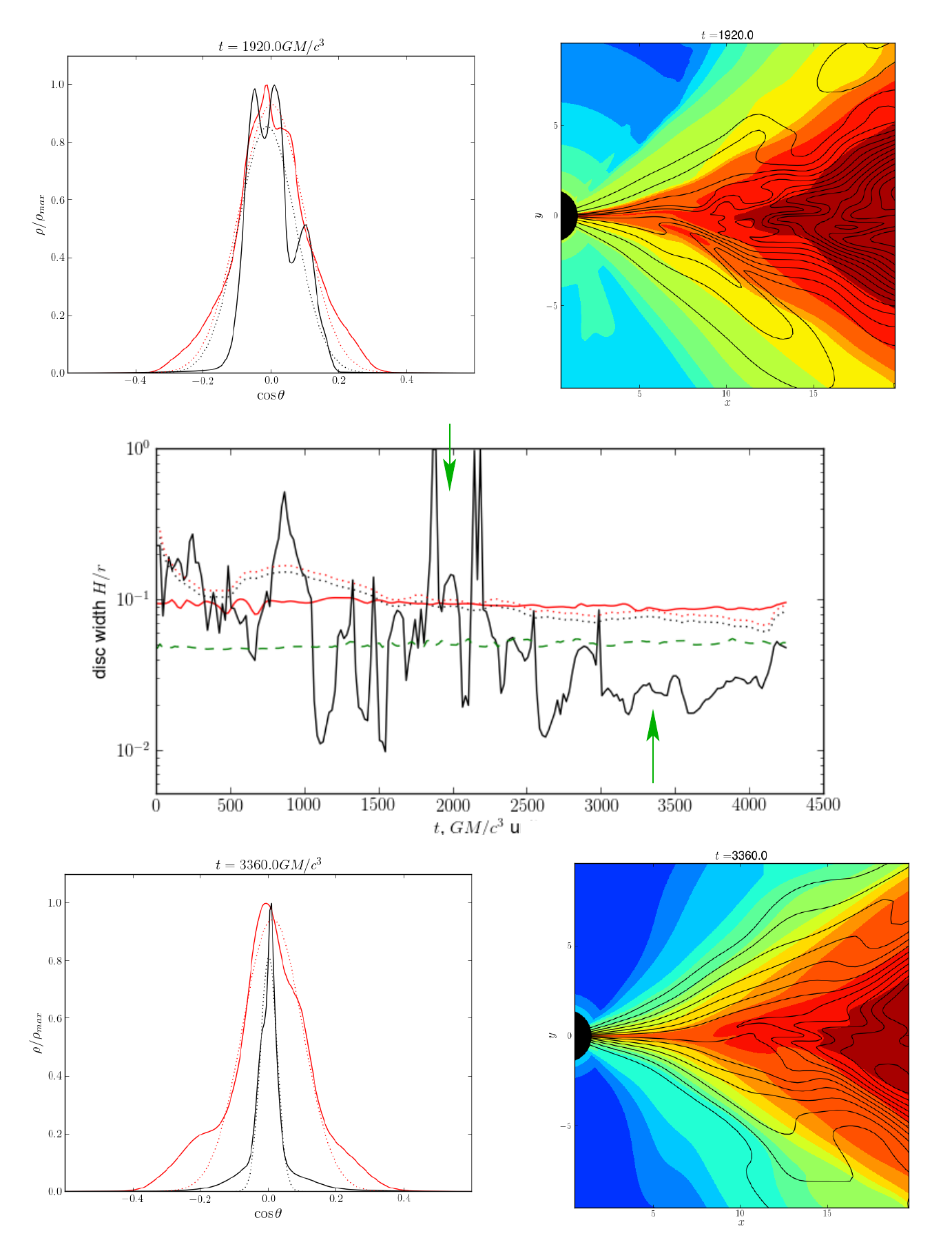}
\caption{ Same as previous figure, but for $a=-0.9$ (counter-rotation, C1 model).
 } 
\label{fig:harm-9}
\end{figure*}

\begin{figure*}
 \centering
\includegraphics[height=0.9\textheight]{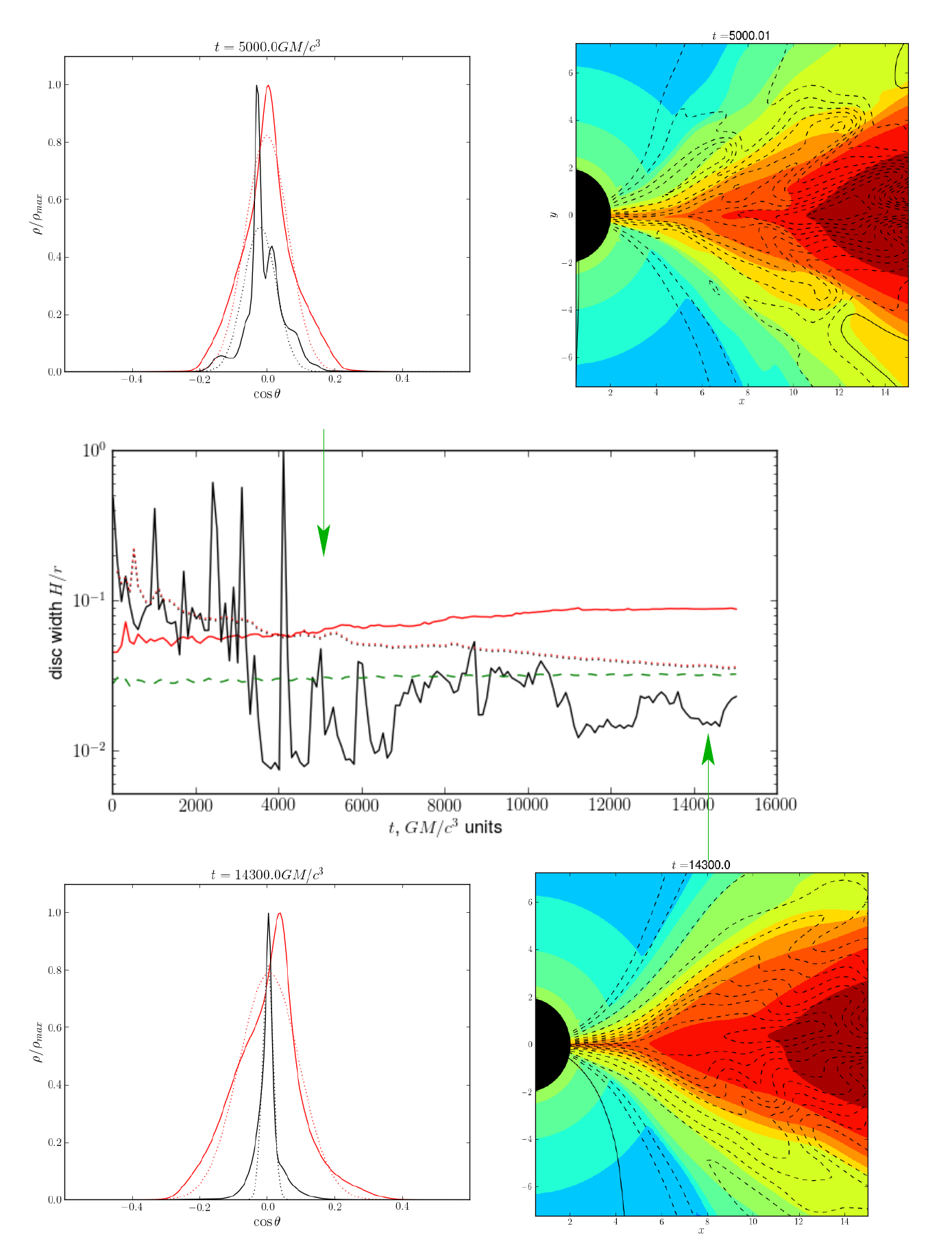}
\caption{ Same as previous figure, but for the A50s model ($a=0$, symmetrical multi-loop
  initial magnetic field).
 } 
\label{fig:harm50s}
\end{figure*}

\begin{figure*}
 \centering
\includegraphics[height=0.9\textheight]{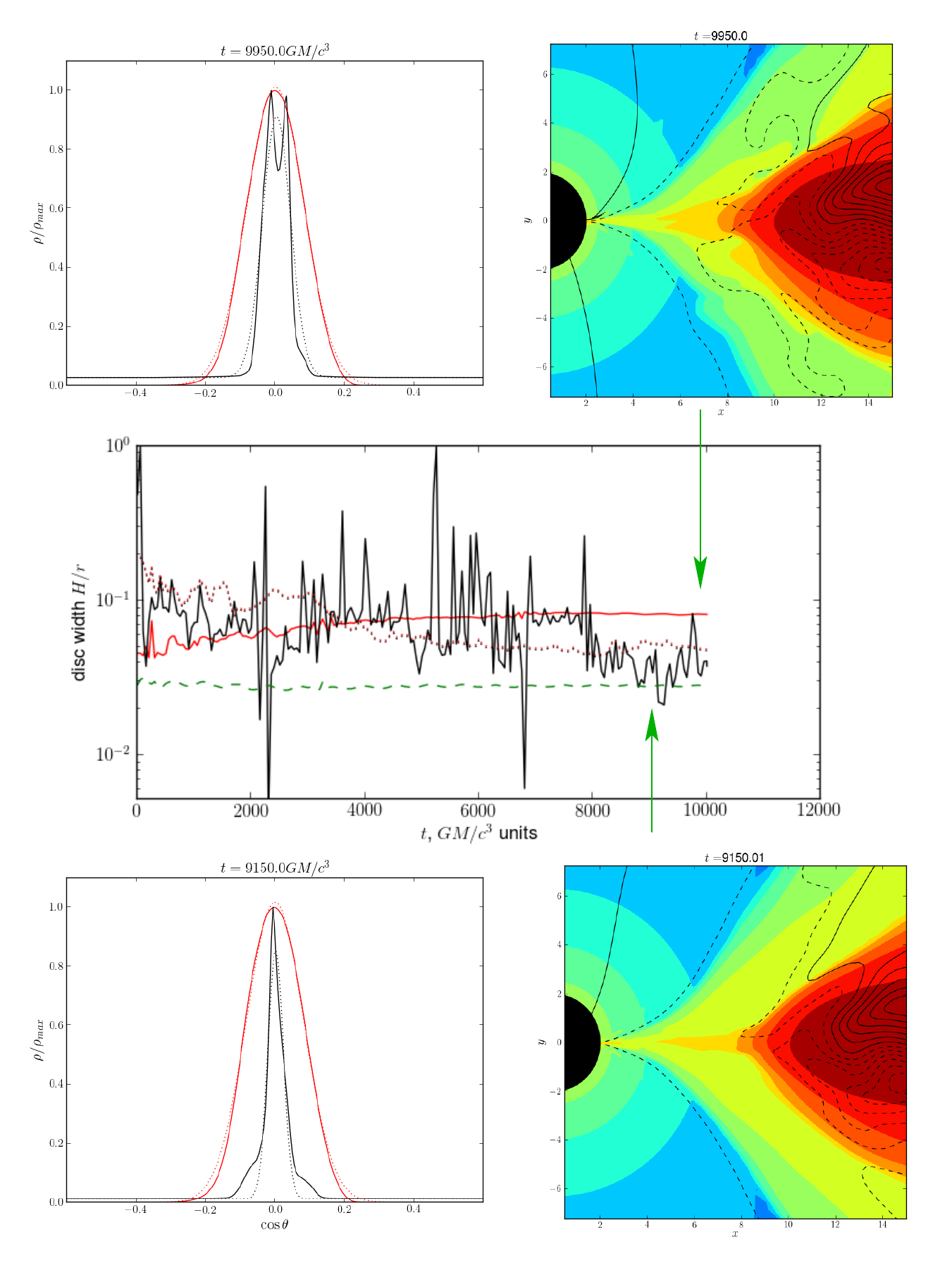}
\caption{ Same as previous figure, but for the A50a model ($a=0$,
  antisymmetric multi-loop initial magnetic field).
 } 
\label{fig:harm50a}
\end{figure*}

\subsubsection{Disc thicknesses}\label{sec:res:htor}


Disc thickness was measured by Gaussian fitting of density distributions
with polar angle, averaged within one of the two radial ranges: 
between one and two ISCO radii and
inside the {\it marginally bound orbit} with $r_{mb} = 2-a +
2\sqrt{1-a}$. Starting with this radius rather than $r_{ISCO}$ allows to ignore the transition region
where initial radial velocity and toroidal magnetic fields are
important. Importance of the marginally bound orbit is clearly seen in
simulations by \citet{abram10} where viscous stresses are still dynamically
important down to $r_{mb}$ in transonic discs but not further. 
Vertical density scale in the first region $(1..2)r_{ISCO}$ 
was taken as an estimate for
$h_0=\left(H/r\right)_0$, while in the second $r_{hor}..r_{mb}$ 
was compared with $h_{eq} = H_{eq}/r$ for the
actual span of radial coordinates. For $h_0$, the Gaussian $\sigma$ for all
the models is of
the order 0.1 that corresponds to half-width thickness of about 0.2. Thus the
disc may be considered reasonably geometrically thin. For the $\alpha$,
$\beta$ and $h_0$ values in the simulations, the inner flow should be thinner
than the disc itself. 

Different vertical scales are compared in figures \ref{fig:harm0},
\ref{fig:harm9} and \ref{fig:harm-9} for the single-loop case and in figures
\ref{fig:harm50s} and \ref{fig:harm50a} for multi-loop
configurations. Individual snapshots are also given as
examples of density and magnetic field behaviour. Vertical contraction is more
profound than one would expect in a magnetically supported disc but is rather
stopped by thermal pressure 
(compare the black solid curve to the green dashed line
in the five figures) or even further. However, B1h model suggests that
spatial scales smaller than $H_m$ may appear only on coarse meshes but the
$H_m$ scale is stable. The $H_{eq}$ spatial scale is in fact present but is
better seen if one considers density distribution averaged in time or if
  magnetic stresses are considered rather than densities. 
In figure \ref{fig:tzz}, vertical structure averaged over time and
radii between $r_{hor}..r_{mb}$ is shown for thermal pressure and vertical
($T_z^z$) magnetic stress component. Magnetic stress seems to vary on spatial
scales close to $H_{eq}$ apart from $H_{m}$. The pressure
maximum also becomes broader after averaging because the positions of the very
narrow ($H\sim 0.01$) maxima in the free-falling flow vary with time. Sometimes,
even two or more such streams are present separated by distances closer either
to the initial disc thickness or to the $H_{eq}$ scale dictated by magnetic
field stresses.

In the multi-loop case, there is as well no general consistence with the estimated flow
vertical scale and predictions of the analytical model. Symmetric and
antisymmetric models behave similarly. 
Finer structure connected to individual loops and
current sheets is seen in individual slices even better (see figures
\ref{fig:harm50s} and \ref{fig:harm50a})

It seems that the MHD flow is in fact unstable to the processes that split the
flow into thin streams with lower magnetization supported by thermal pressure
rather than magnetic fields, sandwiched into a thicker magnetized layer with
the vertical scale of about $H_{eq}$. Since magnetic field changes
direction (see figure \ref{fig:tzz}), the structure of the flow may be
characterised as a driven current
sheet. Reconnections in this current sheet may be one of the sources of energy
released inside the ISCO. The spatial scales of the current sheets, however, 
may be beyond the scope of MHD approach. For example,
the ``thin current sheets'' in terrestrial magnetosphere have thicknesses of
the order of ion gyro-radii \citep{zelenyi}, the spatial scale that MHD
approach can not reveal since it does not account for the masses of the
particles and existence of different particle populations. 

\begin{figure*}
 \centering
\includegraphics[width=1.0\textwidth]{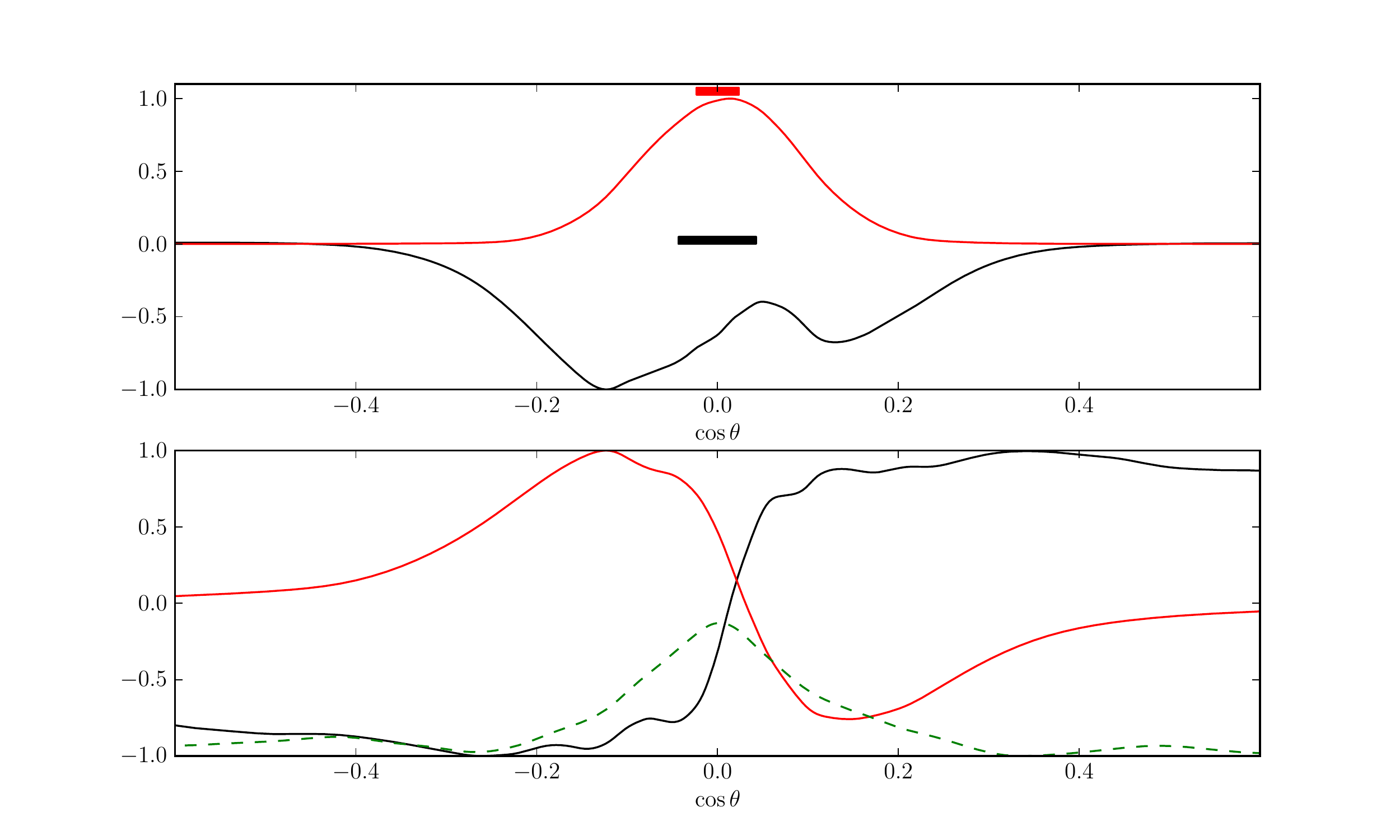}
\caption{Vertical structure of the free-falling flow averaged in time
  (2000..5700) and in the radial range of $r_{hor}..r_{ISCO}$, model B1h. 
  In the upper panel, thermal pressure
  (red/grey) and electromagnetic stress component $-T_z^z$ are shown together
  with the bars representing the estimated thickness scales $H_m/r$ (red/grey)
  and $H_{eq}/r$ (black) at $r_{mb}$. Lower panel shows similarly averaged
  magnetic field components: $B^{\hat{r}}$ (solid black), $B^{z}$ (dashed
  green) and toroidal $B^{\hat{\varphi}}$ (solid red/grey). 
 } 
\label{fig:tzz}
\end{figure*}

{
The observed inflow structure has much in common with the non-linear MRI
solution found by \citet{GX94} for the case of extreme compressibility. 
In this solution, the medium is vertically stratified into regions with
different horizontal field direction divided by thin dense layers of
compressed matter that exists near the nodes of horizontal magnetic
field. 
Such a solution emerges if magnetic field tension becomes 
much stronger than
thermal pressure that is indeed the case for the flow inside the ISCO. 
There is usually only one radial field alternation in the inner flow, probably
because the
vertical scale of a magnetically supported disc essentially coincides by the
order of magnitude with the
wavelength of the fastest-growing MRI mode $\lambda_{MRI} \sim 2\pi
v_{Az}/\Omega_z$. 
}

\section{Discussion}\label{sec:disc}

\subsection{Observability of the inner flow}

 As it was mentioned in the Introduction, the inner flow receives enough
  power to become comparable in its luminosity to the accretion disc
  itself. The question is whether the flow is cooled efficiently enough to
  emit significant part of this power. 
Compton cooling time is likely less than the free-falling time
(see for example \citet{matteo97}):

$$
t_C \sim \dfrac{1}{l_S}\dfrac{GM}{c^3} \sim \dfrac{m_e}{m_p} \dfrac{1}{l_{Edd} \Theta} \times \dfrac{GM}{c^3},
$$

where 
$\Theta = kT/m_ec^2$ and $l_S = \dfrac{r}{L} \dfrac{\sigma_T}{m_ec^3}$ is seed photon
compactness. The above formula assumes $r \sim GM/c^2$ and $L \sim l_{Edd}
\dfrac{4\pi GM m_p}{c\varkappa}$, $l_{Edd}$ is Eddington ratio. 
Dimensionless temperature may be estimated by considering the balance of
heating and cooling mechanisms as $\Theta \sim 0.1 \tau \left(L_H/L_S\right)^{1/4}
\sim 0.1$ (see \citet{PK95}) where $L_{H,S}$ are heating and seed photon
luminosities, and $\tau$ is scattering optical depth in the
inner flow. 

The inner flow is cooled efficiently by inverse Compton losses since its
temperature is about tens of keV and luminosity is determined by the thermal
and possibly magnetic energy stored inside the disc (see Introduction). 
Hence the inner flow is a probable candidate for the source of
X-ray radiation of quasars that is known to emerge from a region of the size
$\sim r_{ISCO}$. The characteristic X-ray luminosities of QSO are about
$0.01-0.1$ of their bolometric luminosities \citep{qsoxrays}, and the X-ray spectrum is
generally consistent with non-saturated Comptonization.  


\subsection{Limitations of the model}

If the inner flow scatters a large number of seed photons (produced in the
disc or near the last stable orbit), it will also experience radiation
drag. I suppose that this effect is essentially unimportant for angular
momentum transport because the angular momenta of matter and radiation leaving
the last stable orbit are more or less the same. However, if rotation
parameter becomes close to the Thorne's critical value of $a\simeq 0.998$
\citep{thorne74}, photons with lower angular momenta will enter the
regions close to the black hole more often. On the other hand, photons with larger
angular momenta may spend more time on grazing orbits. These effects are
probably subtle unless Kerr parameter is high. 


Another possible effect that may alter the radial structure of the inner flow
is angular momentum transfer by non-diagonal terms in electromagnetic
stress-energy tensor. If these terms are important, estimate~(\ref{E:ur})
for the radial velocity 
needs to be replaced by more comprehensive dynamical equations. Besides,
specific energy and angular momentum may now vary with radius. 
Magnetic field will be important in angular momentum transfer if its
contribution to stress-energy tensor becomes comparable to that of in-falling
matter. { Condition that magnetic field effects are {\it small} in the weakly
magnetized flow is:

$$
\dfrac{B^r B^\varphi}{4\pi\rho u^r u^\varphi} < 1,
$$

where $\rho$ is the rest mass density of the (cold, non-relativistic in sense
of its temperature)
in-falling matter. Both velocity components are moderately relativistic $u^r \sim u^\varphi \sim 1$, and
$\rho \sim \dfrac{\dot{M}}{4\pi H r u^r}$. Taking into account equipartition
in the disc (equation~(\ref{E:b0})), magnetic field tension component 
may be roughly estimated as:

$$
\dfrac{B^r B^\varphi}{4\pi} \sim B_0^2 \times \dfrac{1}{4\pi} \left(\dfrac{H_0 r_0}{Hr}\right)^2 \sim
\dfrac{\dot{M}v_K^0}{4\pi \alpha\beta H_0 r_0 } \left(\dfrac{H_0 r_0}{Hr}\right)^2
$$

Hence:

$$
\dfrac{B^r B^\varphi}{4\pi \rho u^r u^\varphi} \sim \dfrac{1}{\alpha\beta} v_K^0 u^r
\dfrac{H_0 r_0}{Hr} \sim \dfrac{H_0 r_0}{Hr} \sim \dfrac{1}{\alpha\beta}
\dfrac{H_0 r_0}{H r}
$$

If $H=H_{eq}$ (equations (\ref{E:eq:polout}-\ref{E:eq:torout})), the above
condition may be re-written as:

\begin{equation}
\left(\dfrac{H_0}{\alpha\beta}\right)^{2/3} \dfrac{r_0}{r} \lesssim 1
\end{equation}

This condition is easily fulfilled for thin discs if $\alpha\beta
\gtrsim 1$ and $H_0 \lesssim 1$. }
{ 
It may still hold even if $\alpha\beta \lesssim 1$ but for accretion discs
thin enough.
The model becomes invalid for $h_0 \gtrsim \alpha \beta$ for one more reason
illustrated by the empirical scalings
(\ref{E:eq:polout}-\ref{E:eq:torout}). If the relative disc thickness is
larger than $\alpha \beta$, the inner flow is geometrically thick and the
expressions for the radial gravity become incorrect. Geometry of the flow may
also change qualitatively if it becomes thicker. In particular, some part of
it may escape to infinity forming a jet or some other type of an outflow. 
}

\subsection{Conclusions}

I conclude that the flow inside the last stable orbit should have a disc-like
structure with the vertical spatial scale $\sim
\left(\dfrac{h_0}{\alpha\beta}\right)^{1/3}$. Numerical simulations support
existence
of similar vertical scales but the flow actually has a more complicated vertical
structure, possibly containing single or multiple thinner layers (or streams, in 3D)
supported by thermal pressure rather than magnetic stresses. 
The reason is probably that the magnetically supported flow is unstable and
the turbulent magnetic field configuration has the tendency to collapse 
into current sheets
with less-magnetized interior 
supported in vertical direction by thermal pressure rather than magnetic
fields. 

\section*{Acknowledgments}

The author would like to thank Alexander Tchekhovskoy for his help with {\it
  HARM}, Kirill Sokolovsky for providing me with additional computational
resources, and the referee (Chris Reynolds) for drawing my attention towards
the MHD instabilities operating in the magnetized flow. 
I also acknowledge support from the Dynasty Foundation and from the
RFBR grant 14-02-91172~GFEN\_a. 

\bibliographystyle{mn2e}
\bibliography{mybib}                                              

\end{document}